\documentclass[pra,twocolumn,showpacs,showkeys,footinbib]{revtex4-1}

\usepackage{graphicx}
\usepackage{amsmath}
\usepackage{amssymb}
\usepackage{mathtools}

\usepackage[english]{babel}
\usepackage{dsfont}
\usepackage{bbm}
\usepackage{hyperref}
\usepackage{amsmath, amsthm, amssymb, mathrsfs}
\usepackage{times}
\usepackage[scriptsize]{subfigure}

\usepackage[usenames,dvipsnames]{color}

\newcommand{\la}{\lambda}

\newcommand{\dg}{\dagger}
\newcommand{\gam}{\gamma}
\newcommand{\LE}{\mathcal{L}}
\newcommand{\pd}{\partial}
\newcommand{\F}{\mathcal{F}}

\newcommand{\sig}[2]{\sigma_{#1}^{#2}}

\def\ket#1{\mathinner{\left|{#1}\right\rangle}}
\def\braket#1{\mathinner{\left\langle{#1}\right\rangle}}

\newcommand{\Braket}[2]{\mathinner{\left\langle\left.#1\right\vert #2 \right\rangle}}
\def\ketbra#1{\mathinner{\left|{#1}\right\rangle}\mathinner{\left\langle{#1}\right|}}

\def\bravert{\egroup\,\vrule\,\bgroup}

\begin{document}
\title{Phase diagram and quench dynamics of the Cluster-\textit{XY} spin chain}

\author{Sebasti{\' a}n Montes}
\affiliation{Perimeter Institute for Theoretical Physics,
31 Caroline St. N, N2L 2Y5, Waterloo Ontario, Canada}
\affiliation{University of Waterloo, 200 University Ave W, Waterloo ON, N2L 3G1, Canada }
\email{smontesvalencia@perimeterinstitute.ca}

\author{Alioscia Hamma}
\affiliation{Perimeter Institute for Theoretical Physics, 31 Caroline St. N, N2L 2Y5, Waterloo ON, Canada}
\email{ahamma@perimeterinstitute.ca}

\begin{abstract}
We study the complete phase space and the quench dynamics of an exactly solvable spin chain, the Cluster-\textit{XY} model. In this chain, the Cluster term and the \textit{XY} couplings compete to give a rich phase diagram. The phase diagram is studied by means of the quantum geometric tensor. We study the time evolution of the system after a critical quantum quench using the Loschmidt echo. The structure of the revivals after critical quantum quenches presents a non trivial behavior depending on the phase of the initial state and the critical point. 

\end{abstract}


\maketitle

In recent years, the understanding of quantum phases and quantum phases transitions has gained an incredible momentum. This is due 
to both strong theoretical advance, stimulated by the richness of quantum phases, and experimental advance. On the theoretical side, there has been great progress in understanding how 
quantum critical points affect the finite temperature regime \cite{Sachdev-QPT} and quantum entanglement\cite{amico}. Moreover, we are realizing that quantum phases host rich novel quantum orders and phases of matter \cite{wenbook}. 
From the experimental side, ultra cold atom gases have proven to be the ideal arena to see coherent quantum evolution for many body systems  \cite{Experiments cold atoms 1, Experiments cold atoms 2, Cold atoms}.  

The study of closed quantum systems out of equilibrium is important for manifold reasons. To start with, one is interested in applications to quantum information, in which decoherence and entanglement dynamics play a fundamental role. An important theoretical perspective regards the understanding of the notion of universality for a system away from equilibrium, where the traditional concepts of phase, renormalization group and fixed point fail. Recently, the study of the equilibration of quantum many body systems has given new insights in the foundations  of statistical mechanics \cite{popescunature, popescupre, Non-equilibrium review, Quantum quenching}.

Driving a system out of equilibrium can be accomplished in  many ways. Most of the efforts have focused on \textit{quantum quenches} \cite{Quantum quenching}, namely sudden global or local changes of the external parameters of the Hamiltonian governing the unitary evolution of the closed system. One of the ways of understanding the dynamics of a system after a quench is the \textit{Loschmidt echo}, a measure of the partial recurrences with the original state as a function of time \cite{Loschmidt - Review, Loschmidt - Review 2}. 
Recently, the time behaviour of the Loschmidt echo has been investigated in  various models, in particular the \textit{XY} spin chain \cite{Zanardi - Unitary equilibration Loschmidt, Zanardi - Statistics Loschmidt echo, Zanardi - Decay Loschmidt, RevivalLiebRobinson}.

In this paper we will consider a one-dimensional model that extends the \textit{XY} spin chain with a three-body cluster term. The exact solution becomes available using well-known techniques and this allows us to study the complete phase diagram. We find that a particular critical region has a behavior quite different to the one found in the \textit{XY} model. We then study the behavior of the Loschmidt echo after critical quenches for two different critical points. We find qualitative differences derived from the non trivial nature of the phase space. 


\textit{Cluster-\textit{XY} spin chain.---} Cluster states have emerged recently as a physical system for implementing one-way quantum computation. In particular, it has been shown that two-dimensional cluster states serve as fiducial states for universal measurement-based quantum computation  \cite{MBQC-Original, b3, MBQC-Nature, Wei, Miyake}.

The cluster state can be defined using a so-called stabilizer Hamiltonian. Consider a finite-dimensional lattice $\mathbb{L}$ composed of $N$ vertices, each vertex containing a two-dimensional quantum system (qubit). The cluster state for this system can be defined as the unique $+1$ eigenstate of the stabilizer operators
\begin{equation}
 K_\mu = \sig{\mu}{z}\prod_{\nu\sim\mu}\sig{\nu}{x},\qquad \mu,\nu\in\mathbb{L},
\label{Cluster term}
\end{equation}
where $\nu\sim\mu$ denotes that $\nu$ is connected to $\mu$ and $\sigma^\alpha$ are the Pauli matrices \cite{Bartlett-Phase transitions}. The stabilizer Hamiltonian is simply
\[H_C = -\sum_{\mu\in\mathbb{L}}K_\mu,\]
and the cluster state is defined as its ground state. This preparation can also be achieved by preparing all the qubits in the $\ket{0}$ state ($\sigma^z\ket{0}=\ket{0}$) and then performing a controlled sign operator $U=\exp(i\pi\ketbra{+}\otimes\ketbra{+})$ (where $\sigma^x\ket{+}=\ket{+}$) on every pair of connected vertices \cite{Bartlett - Identifying phases}.
The model we  study incorporates the one-dimensional version of the cluster phase competing with the \textit{XY} model in a transverse field. The Hamiltonian is
\begin{align}
 H=&-\sum_{i=1}^N \sig{i-1}{x}\sig{i}{z}\sig{i+1}{x}-h\sum_{i=1}^N\sig{i}{z}\nonumber\\
&+\la_y\sum_{i=1}^N \sig{i}{y}\sig{i+1}{y}+\la_x\sum_{i=1}^N \sig{i}{x}\sig{i+1}{x},
\label{ClusterIsing Hamiltonian}
\end{align}
where $\sig{n}{\alpha}$ ($\alpha=x,y,z$) are the Pauli matrices acting on the site $n$ of the lattice and we impose periodic boundary conditions ($\sig{N+1}{\alpha}\equiv\sig{1}{\alpha}$).  Similar models were considered in \cite{Bartlett-Phase transitions,Bartlett - Identifying phases,ClusterAFM,Stat Mech Cluster Ising}. Defining local raising and lowering operators $\sig{n}{\pm}=\frac{1}{2}(\sig{n}{x}\pm i\sig{n}{y})$, we obtain global canonical anticommutation relations by using a Jordan-Wigner transformation \cite{Sachdev-QPT}
\begin{equation}
 c_l^\dg=\left(\prod_{m=1}^{l-1}\sig{m}{z}\right)\sig{l}{+},
\end{equation}
so that the model is mapped to a quadratic Hamiltonian of spinless fermions $\{c_n,c_m\}=0,\, \{c_n,c_m^\dg\}=\delta_{nm}$. Note that the parity operator $Q:=\prod_{n}\sig{n}{z}$ commutes with the Hamiltonian and can be diagonalized simultaneously with it. Using the fact that the system has translational invariance, we may perform a Fourier transform
\begin{equation*}
c_k=\frac{1}{\sqrt N}\sum_{n=1}^N e^{ikn}c_n,\quad k=\frac{\pi}{N}(2m+1-q),
\label{Fourier transform}
\end{equation*}
where we decompose the Hilbert space so that $Q=(-1)^q $ and $ m=0,...,N-1$. We can then rewrite the Hamiltonian as
\begin{equation*}
 H=2\sum_{0\leq k\leq\pi}\left[\epsilon_k(c_k^\dg c_k+c_{-k}^\dg c_{-k})+i\delta_k(c_k^\dg c_{-k}^\dg+c_k c_{-k})\right],
\end{equation*}
up to a constant, where
\begin{subequations}
\begin{align}
\epsilon_k&=\cos(2k)-(\la_x+\la_y)\cos(k)-h,\\
\delta_k&=\sin(2k)-(\la_x-\la_y)\sin(k).
\end{align}
\end{subequations}
 We diagonalize the Hamiltonian by means of a Bogoliubov transformation
\begin{equation}
 \gam_k=\cos(\theta_k/2)c_k-i\sin(\theta_k/2)c_{-k}^\dg,
\label{Bogoliubov transformation}
\end{equation}
imposing $\theta_{-k}=-\theta_k$ so that $\{\gam_k,\gam_{k'}^\dg\}=\delta_{kk'}$. The Hamiltonian becomes
\begin{equation}
 H=2\sum_{0\leq k\leq\pi}\Delta_k\left(\gam_k^\dg\gam_k+\gam_{-k}^\dg\gam_{-k}-1\right)+\text{const.,}
\label{Diagonal Hamiltonian}
\end{equation}
where we defined the energy for the so called \emph{Bogoliubov quasiparticles} $\Delta_k = \sqrt{\epsilon_k^2+\delta_k^2}$ and we canceled the unwanted $\gam\gam$ terms by choosing $\epsilon_k\sin\theta_k+\delta_k\cos\theta_k=0$, or equivalently
\begin{equation}
 \theta_k=-\arctan\left(\frac{\delta_k}{\epsilon_k}\right).
\label{Bogoliubov angle}
\end{equation}
The ground state has the form of a BCS state in terms of the original operators
\begin{equation}
\ket{\Omega}= \prod_{0\leq k\leq\pi}\left(\cos(\theta_k/2)+i\sin(\theta_k/2)c_k^\dg c_{-k}^\dg\right)\ket{0}_c,
\label{BCS state}
\end{equation}
where $c_k\ket{0}_c=0,$ $\forall k$.

\begin{figure}[h]
  \centering
\subfigure[]{
  \includegraphics[width=0.47\linewidth]{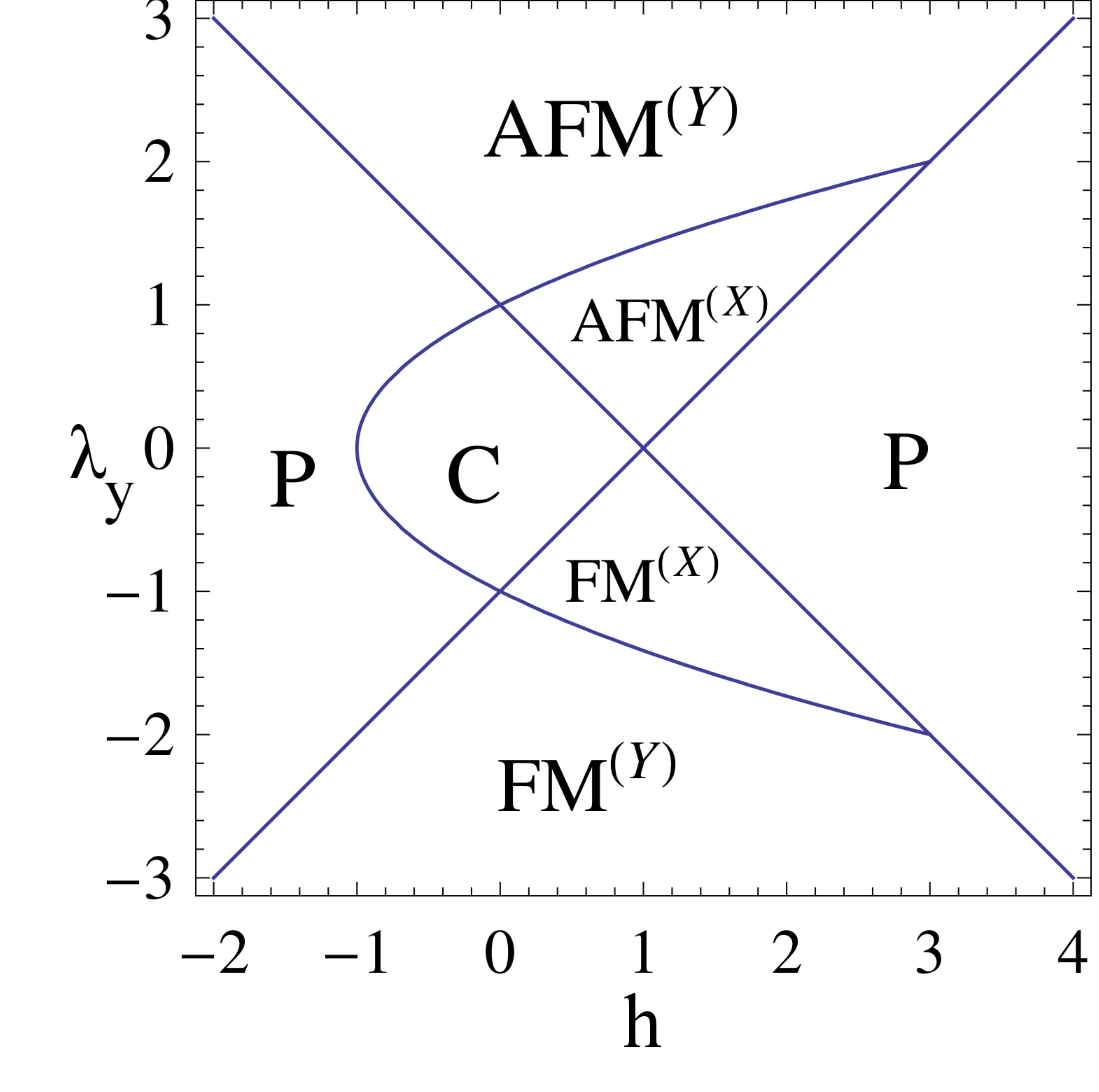}
\label{x0}
}
\subfigure[]{
  \includegraphics[width=0.47\linewidth]{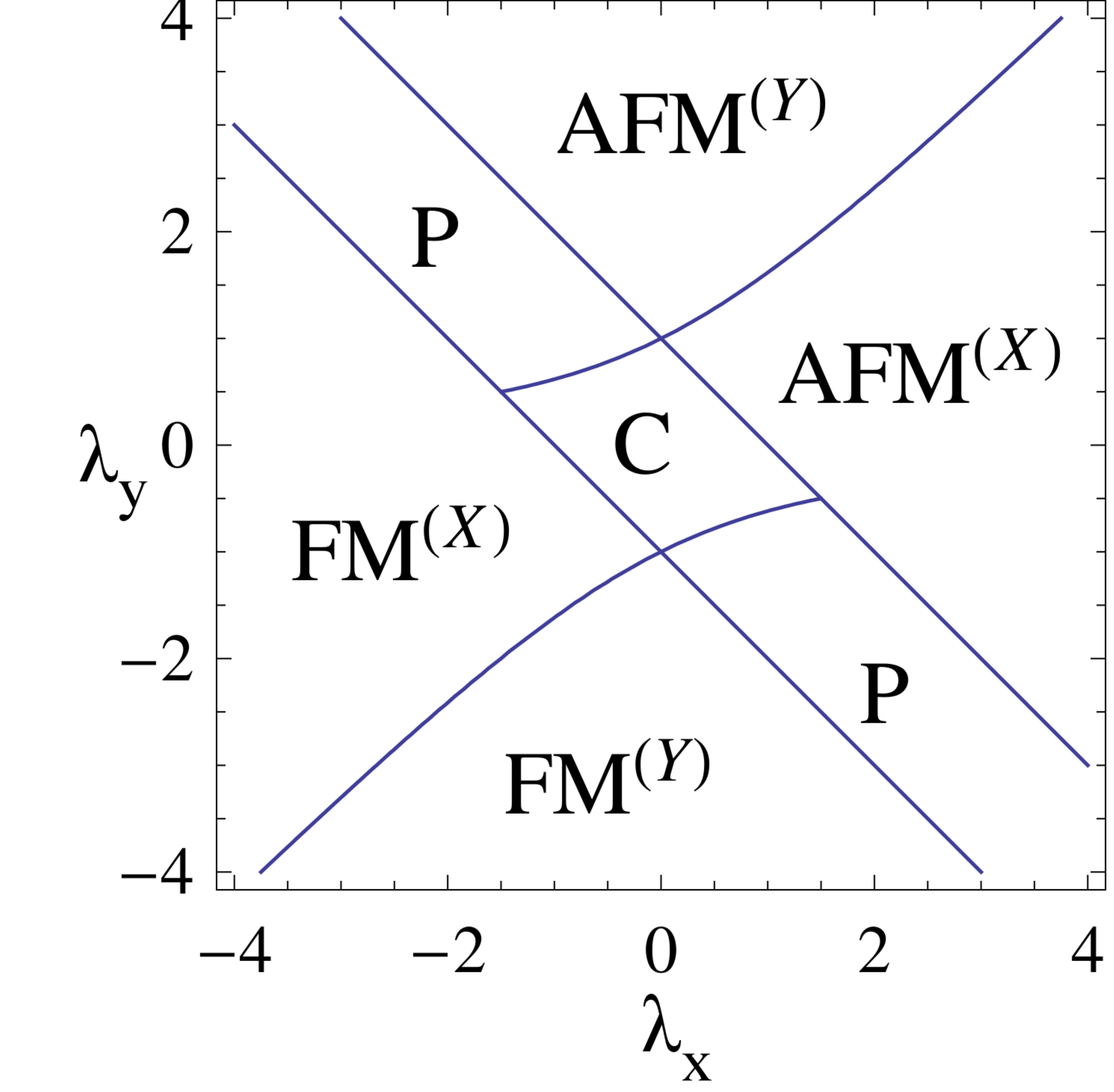}
\label{h0}
}
  \caption{Reduced phase diagrams for (a) $\la_x=0$ and (b) $h=0$. We use the conventions: P (polarized, given by $\sum\sigma^z$), C (cluster), AFM (antiferromagnetic), FM (ferromagnetic).}
\end{figure}

\textit{Phase diagram.---}
At this point, we can draw the phase diagram by finding the regions of quantum criticality where the system becomes \emph{gapless} in the thermodynamics limit, that is, 
 $\Delta_k=0$ for some $k\in [-\pi,\pi)$ when $N\rightarrow\infty$.  First, note that trivially $\delta_k=0$ for $k=0,\pm\pi$.  In that case, $\epsilon_k$ vanishes for
\begin{equation}
 h=\pm(\la_x+\la_y)+1.
\label{Cluster - Gapless 1}
\end{equation}
Now, if $k\neq 0,\pi$, $\delta_k$ vanishes if $\cos(k)=\frac{\la_x-\la_y}{2}$. Using this relation for $\epsilon_k$, we get the critical region
\begin{equation}
 h=\la_y^2-\la_x\la_y-1,\qquad -2\leq \la_x-\la_y\leq 2.
\label{Cluster - Gapless 2}
\end{equation}
We see an asymmetry between $\la_x$ and $\la_y$. This should be no surprise as this is already evident in the cluster term of the Hamiltonian. Note also that the union of the critical regions is invariant under $\la_y\mapsto-\la_y$, $\la_x\mapsto-\la_x$. We can understand this by noting that both local unitary transformations
\[U_1:\quad \sig{2n}{x}\mapsto-\sig{2n}{x},\quad \sig{2n}{y}\mapsto-\sig{2n}{y},\quad \sig{2n}{z}\mapsto\sig{2n}{z} \]
acting only on even sites and
\[U_2:\, \sig{2n+1}{x}\mapsto-\sig{2n+1}{x},\, \sig{2n+1}{y}\mapsto-\sig{2n+1}{y},\, \sig{2n+1}{z}\mapsto\sig{2n+1}{z} \]
acting only on odd sites map $H(\la_x,\la_y,h)$ to $H(-\la_x,-\la_y,h)$. This is a consequence of the $\mathbb{Z}_2\times\mathbb{Z}_2$ symmetry of the cluster state implemented precisely by $U_1$ and $U_2$ \cite{Stat Mech Cluster Ising}.

One of the interesting features of this model is the existence of phases that appear because of the competition between the \textit{XY} and cluster terms. Consider first $\la_x=0$ (Fig. \ref{x0}). The Hamiltonian in this case does not have Ising interactions of the type $\sig{n}{x}\sig{n+1}{x}$. However, two of the regions next to the cluster phase can be connected adiabatically to ferromagnetic and antiferromagnetic states in the $x$ direction, respectively.  Something similar happens for $h=0$. The Hamiltonian does not have a transverse term that tries to polarize all the spins in the same $z$ direction, nevertheless this phase is present in the reduced phase diagram (Fig. \ref{h0}).

\textit{Fidelity susceptibility.---} The phase diagram can be studied \cite{Zanardi - Geometry} by considering the {\em fidelity susceptibility} introduced in \cite{Zanardi - Ground State}, namely the response of the ground state to small changes of the external parameters. Consider a many-body system described by a Hamiltonian
\begin{equation}
 H(\la) = H_0 + \la H_I,
\end{equation}
where $\la$ is an external parameter used to control the system. There is no loss of generality if we write the Hamiltonian this way, especially if the system is large enough so that the critical point is well localized. $H_I$ is called the \emph{driving Hamiltonian}. We may now diagonalize the system and obtain both the energy spectrum and the eigenstates
\begin{equation}
 H(\la)\ket{n(\la)}=E_n(\la)\ket{n(\la)}.
\end{equation}
If we change $\la$ to $\la+\delta\la$ and we are away from possible critical points, the physics described by the neighboring ground states will be similar. However, note that in the thermodynamic limit different ground states will become orthogonal, as was realized by Anderson in the so-called ``orthogonality catastrophe" \cite{Anderson-OC}. For finite size systems, we expect that the new ground state will remain ``close'' to the original ground state and we may study how fast does the overlap goes to zero.  In order to quantify this notion we use the \emph{fidelity} \cite{Zanardi - Ground State,Fidelity Gu}
\begin{equation}
\F(\la,\la'):= |\Braket{\Omega(\la)}{\Omega(\la')}|,
\label{Fidelity}
\end{equation}
where $\ket{\Omega(\la)}$ represents the ground state of $H(\la)$. The response of the fidelity after an infinitesimal change of the external parameter up to second order
reads
\begin{equation}
 \F(\la,\la+\delta\la)=1-\frac{\delta\la^2}{2}\chi_F+\mathcal{O}(\delta\la^4),
\end{equation}
where  the \emph{fidelity susceptibility} \cite{Zanardi-Tensor,Fidelity Gu,Gritsev-Understanding} is defined by
\begin{align}
 \chi_F(\la) := &\Braket{\pd_\la\Omega(\la)}{\pd_\la\Omega(\la)}\nonumber\\
&-\Braket{\pd_\la\Omega(\la)}{\Omega(\la)}\Braket{\Omega(\la)}{\pd_\la\Omega(\la)}.
\end{align}
If we have more than one external parameter, we may generalize this result and obtain the so called \emph{quantum geometric tensor} \cite{Zanardi-Tensor}
\begin{align}
 T_{ab}:=& \Braket{\partial_{\la_a}\Omega(\la)}{\partial_{\la_b}\Omega(\la)}\nonumber\\
 &- \Braket{\partial_{\la_a}\Omega(\la)}{\Omega(\la)}\mathinner{\left\langle\Omega(\la)\left\vert \partial_{\la_b}\Omega(\la)\right.\right\rangle}.
\label{quantum geometric tensor}
\end{align}
In general, this tensor will be complex. Both the real and imaginary parts have nice physical interpretations \cite{Zanardi - Geometry}. The real part will be an induced Riemannian metric on the manifold of parameters
\begin{equation}
 g_{ab}:=\Re \left(T_{ab}\right).
\end{equation}
The geodesics with respect to this metric give information about the optimal adiabatic path that connects to points inside the same quantum phase. Also, the scalar curvature may be used to distinguish different phases and characterize the behavior of critical regions \cite{Zanardi - Geometry}.
The imaginary part is related to the appearance of geometrical phases through the \emph{Berry curvature}
\begin{align}
 F_{ab}&:=\Im\left(T_{ab}\right)\nonumber\\
&=\Braket{\partial_{\la_a}\Omega(\la)}{\partial_{\la_b}\Omega(\la)}-\Braket{\partial_{\la_b}\Omega(\la)}{\partial_{\la_a}\Omega(\la)}\nonumber\\
&=\pd_a A_b-\pd_b A_a,
\end{align}
where $A:=\Braket{\Omega}{\partial_{\la_b}\Omega}$ is the \emph{adiabatic Berry connection} \cite{Hamma - Berry,Zanardi-Tensor}.

\begin{figure}[h]
  \centering
\subfigure[]{
\includegraphics[width=0.45\linewidth]{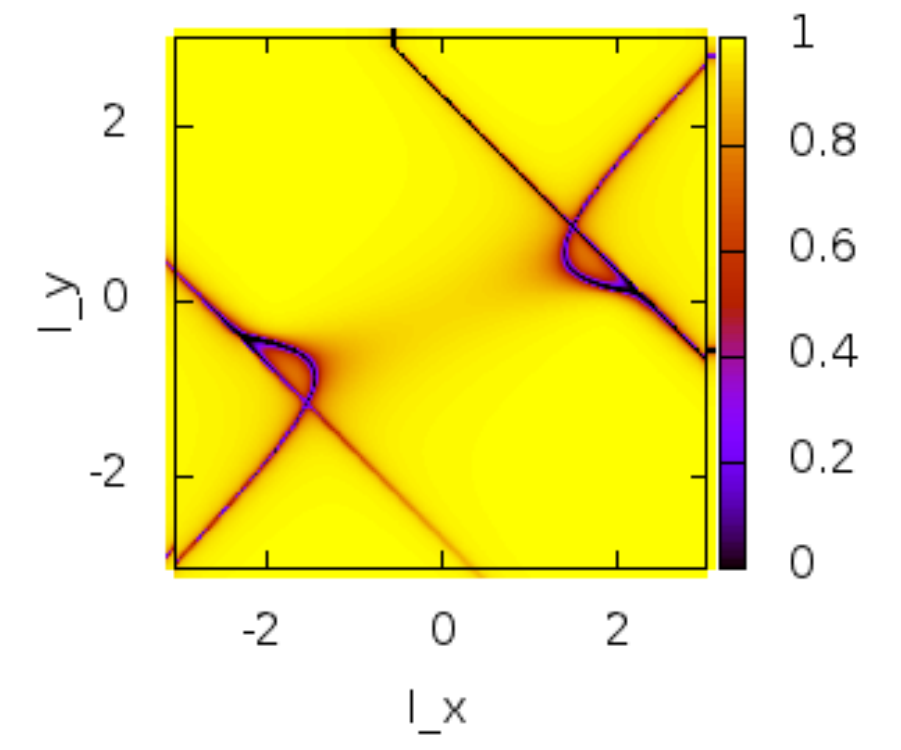}
}
\subfigure[]{
    \includegraphics[width=0.45\linewidth]{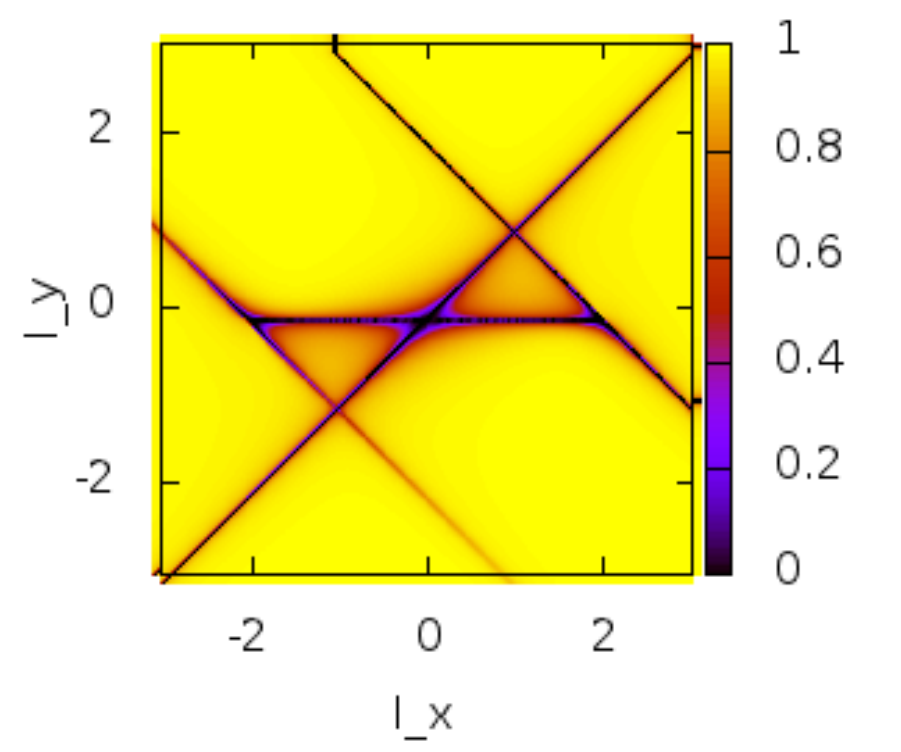}
}
\subfigure[]{
    \includegraphics[width=0.45\linewidth]{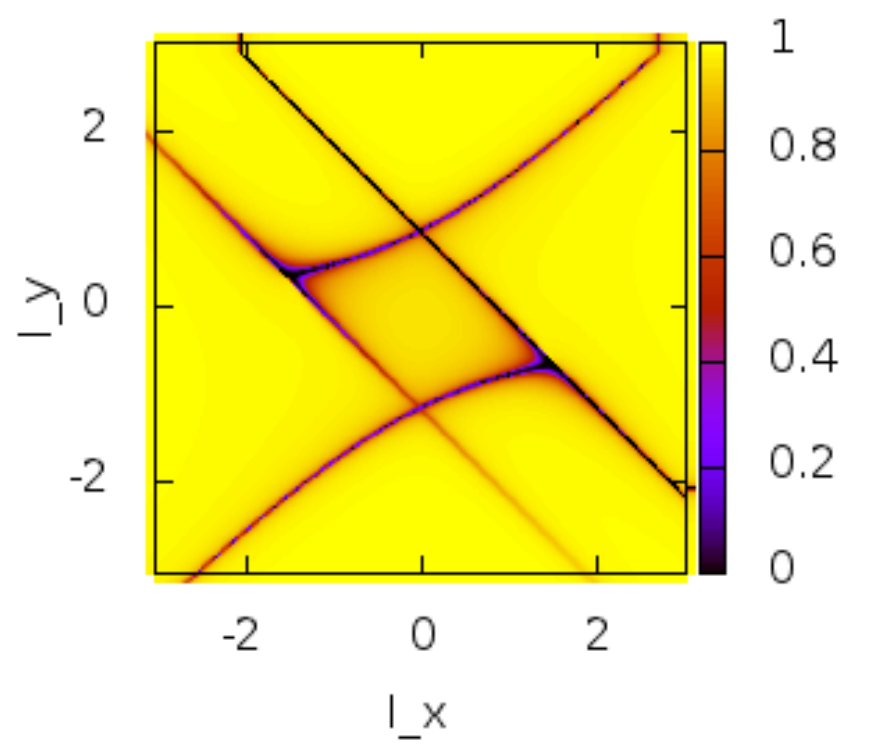}
\label{h0 phase diagram}
}
\subfigure[]{
    \includegraphics[width=0.45\linewidth]{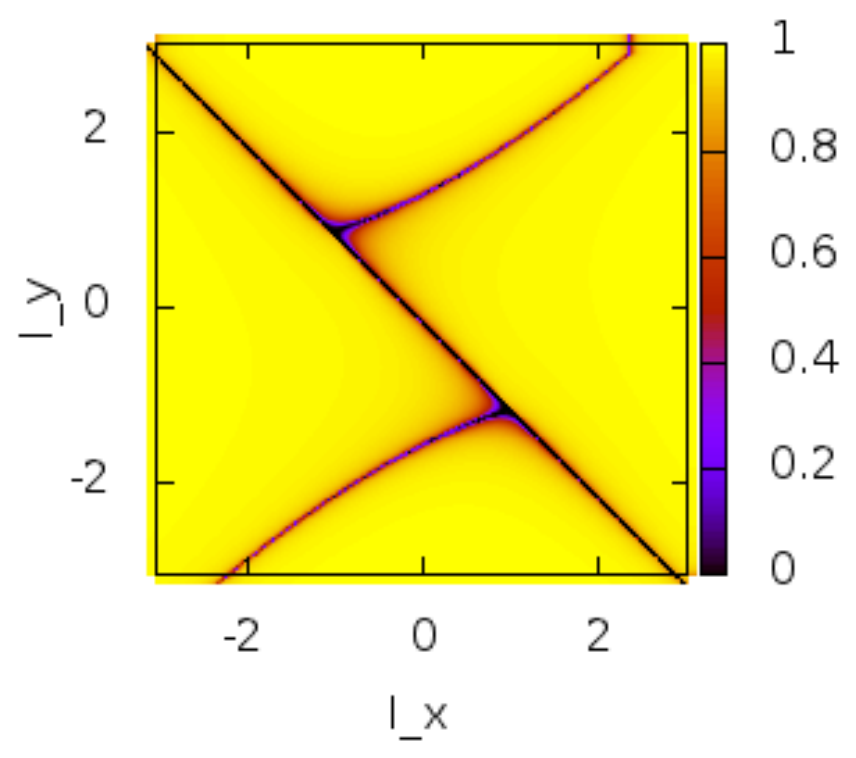}
}
\subfigure[]{
    \includegraphics[width=0.45\linewidth]{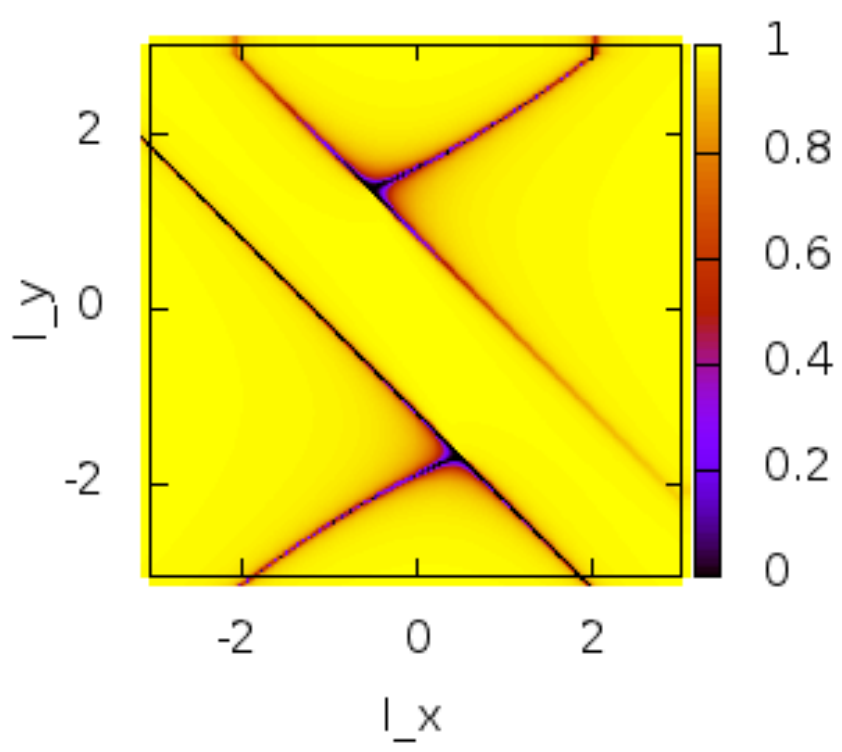}
}
\subfigure[]{
  \includegraphics[width=0.45\linewidth]{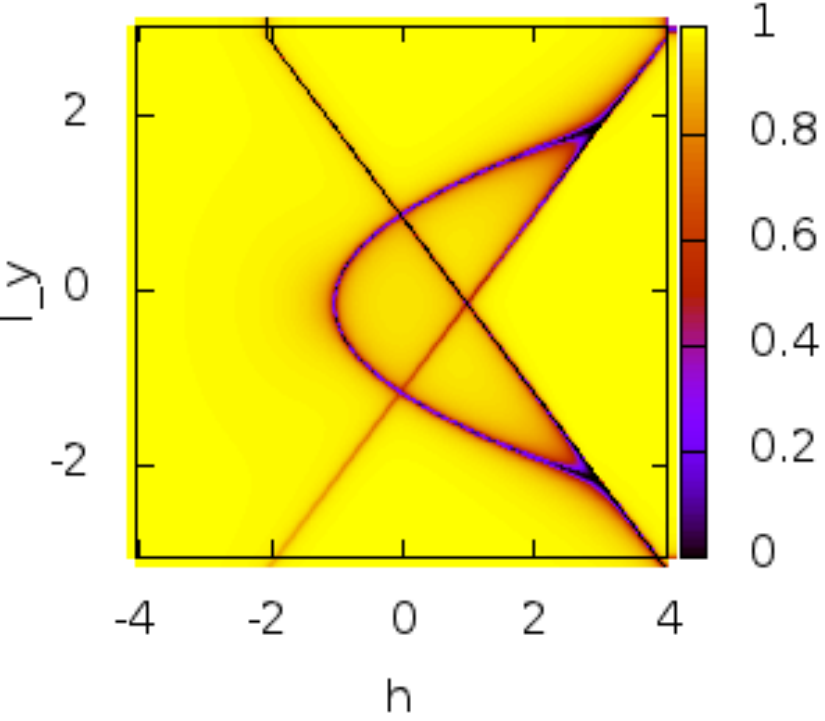}
\label{fid-x0}
}
    \caption{(Color online) Contour plot of the fidelity $\F(\la_y,\la_y+\delta\la_y)$, $\delta\la_y=0.05$, $N=500$, for constant $h$: (a) $h=-1.5$, (b) $h=-1$, (c) $h=0$, (d) $h=1$ and (e) $h=2$. (f) Constant $\la_x=0$ }
    \label{fid plot}
\end{figure}

Coming back to the cluster-Ising model, we may compute the fidelity between neighboring ground states $\ket{\Omega(\la_i)}$  \eqref{BCS state} for different values of the $\{\la_i\}$. If we change  the external parameters $\la_i^{(1)}\rightarrow \la_i^{(2)}$, we can express the ``old'' ground state $\ket{\Omega(\la_i^{(1)})}$ in terms of the operators that diagonalize the ``new'' Hamiltonian $H(\la_i^{(2)})$. The form of the wave function remains the same
\begin{equation}
 \ket{\Omega(\la_i^{(1)})}=\prod_{k>0}\left(\cos\left(\frac{\chi_k}{2}\right)+i\sin\left(\frac{\chi_k}{2}\right)\gam_k^\dg \gam_{-k}^\dg\right)\ket{\Omega(\la_i^{(2)})},
\end{equation}
where $\gam_k,\gam_k^\dg$ are the fermionic operators that diagonalize the Hamiltonian $H(\la^{(2)}_i)$ and we defined
\begin{equation}
 \chi_k=\theta_k(\la_i^{(1)})-\theta_k(\la_i^{(2)}).
\end{equation}

After a straightforward calculation, we obtain
\[\F(\la_x,\la_y,h;\la_x',\la_y',h')=\prod_{0\leq k\leq\pi}\left|\cos\left(\frac{\theta_k-\theta_k'}{2}\right)\right|.\]
From this expression, we compute the quantum geometric tensor  \cite{Zanardi-Tensor}
\[T_{ab}=\sum_{0\leq k\leq\pi}\frac{1}{4}\frac{\pd\theta_k}{\pd\la_a}\frac{\pd\theta_k}{\pd\la_b}. \]
using the convention $\la_1=\la_x$, $\la_2=\la_y$ and $\la_3=h$, where
\begin{align*}
 \frac{\pd\theta_k}{\pd\la_x}&=-\frac{\cos(k)\delta_k-\sin(k)\epsilon_k}{\Delta_k^2},\\
\frac{\pd\theta_k}{\pd\la_y}&=-\frac{\cos(k)\delta_k+\sin(k)\epsilon_k}{\Delta_k^2},\qquad\frac{\pd\theta_k}{\pd h}=-\frac{\delta_k}{\Delta_k^2}. 
\end{align*}
Note that $T_{ab}$ may not be analytic when the system becomes gapless, \textit{i.e.}, when $\Delta_k\rightarrow 0$ for some $k$. Since $T_{ab}$ is a real tensor, this system will have a trivial Berry curvature. The above expressions for the quantum geometric tensor are non trivial. We expect to find a richness of features in their scaling behavior \cite{Zanardi-Tensor}, which can be potentially of use in the optimization of quantum adiabatic algorithms \cite{brac}. A thorough study of the scaling of the geometric tensor in the Cluster-\textit{XY} model is to be found in \cite{next}.

By computing the fidelity in the Cluster-\textit{XY} model, we find the expected critical lines. We illustrate this in figure \ref{fid plot} plotting the phase diagram region that we already discussed in the context of the exact solution.

\begin{figure}[]
  \centering
\subfigure[]{
  \includegraphics[width=0.45\linewidth]{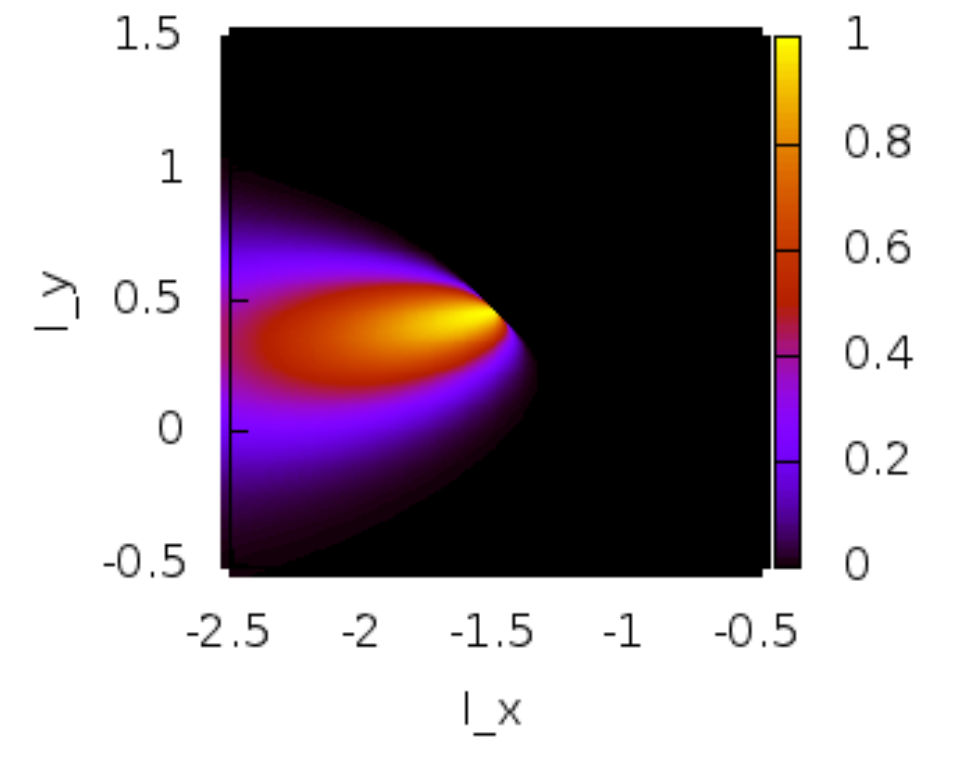}
\label{P1-gs}
}
\subfigure[]{
  \includegraphics[width=0.45\linewidth]{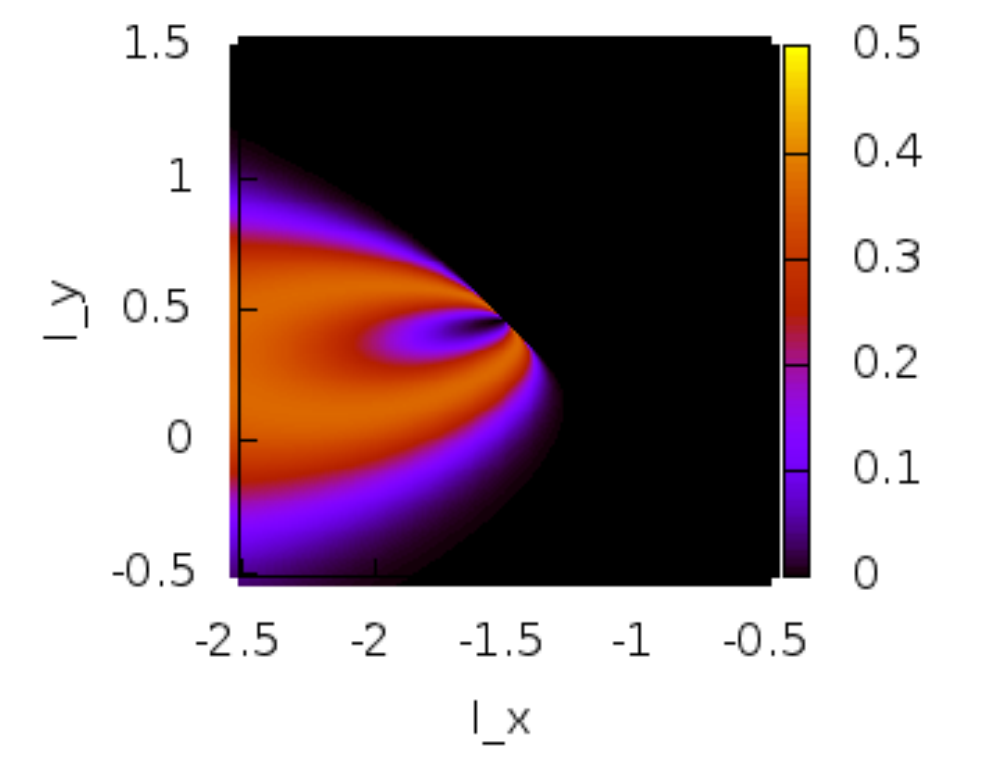}
\label{P1-es}
}
  \caption{(Color online) (a) Overlap between $\ket{\Omega(P1)}$ and the neighboring ground states ($N=500$, $q=1$ and $h=0$). (b) Overlap $F_1(\la)$ between  the two particle states of $P1$ and the neighboring ground states ($N=500$, $q=1$ and $h=0$).}
\label{h0-overlap}
\end{figure}

\begin{figure}[]
  \centering
\subfigure[]{
  \includegraphics[width=0.45\linewidth]{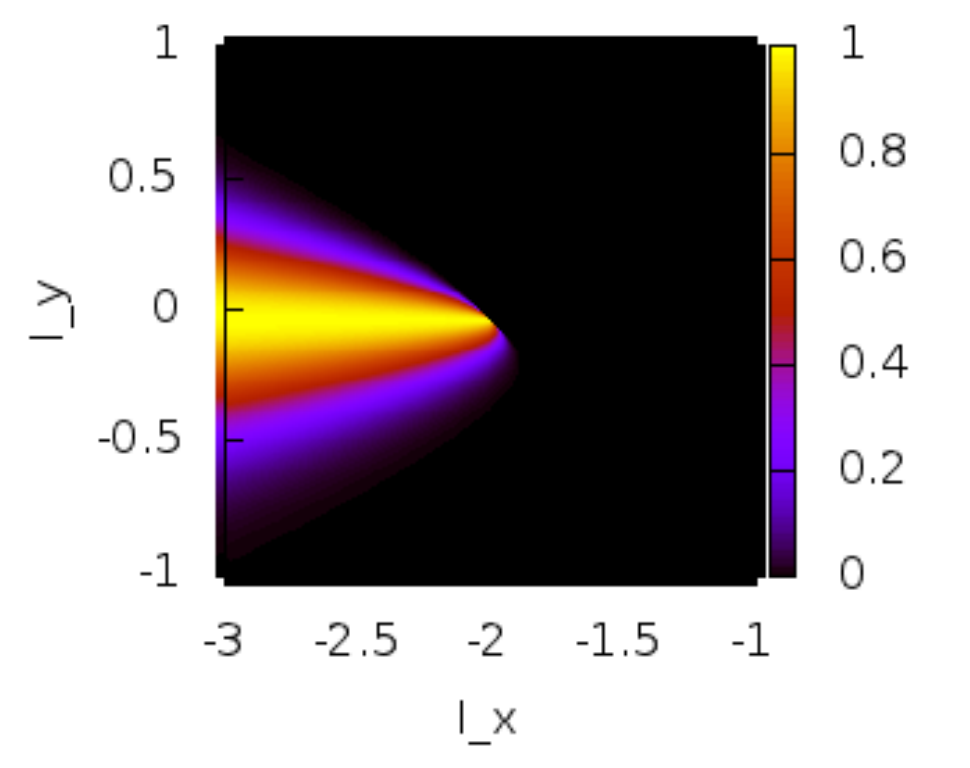}
\label{P2-gs}
}
\subfigure[]{
  \includegraphics[width=0.45\linewidth]{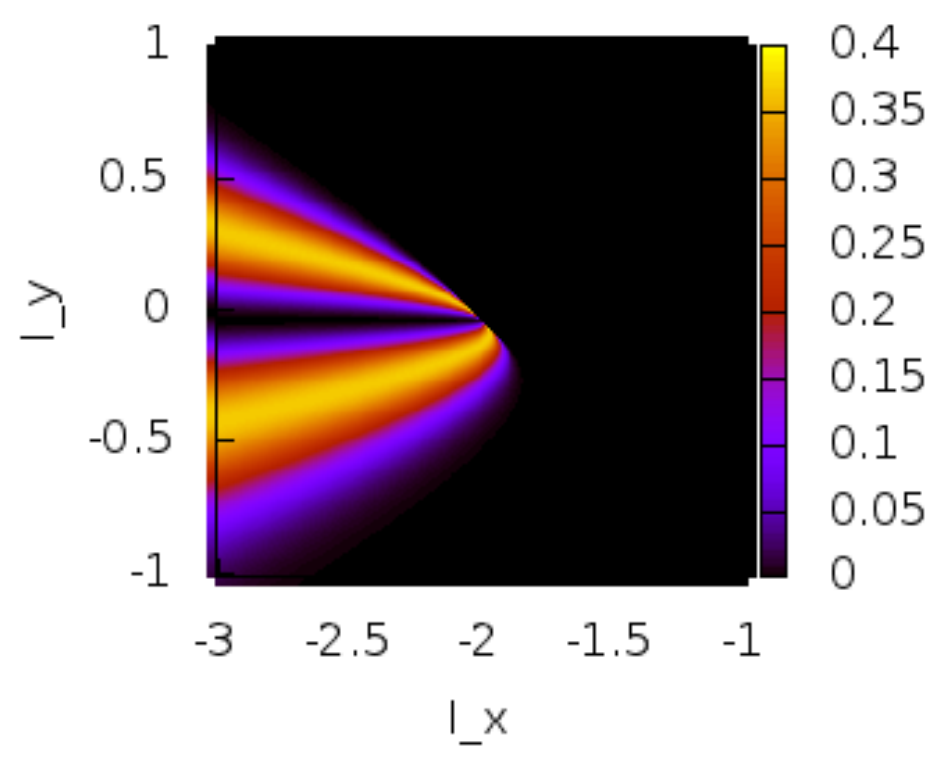}
\label{P2-es}
}
 \caption{(Color online) (a) Overlap between $\ket{\Omega(P2)}$ and the neighboring ground states ($N=500$, $q=1$ and $h=-1$). (b) Overlap $F_1(\la)$ between  the two particle states of $P2$ and the neighboring ground states ($N=500$, $q=1$ and $h=-1$).}
\label{h-1-overlap}
\end{figure}

We notice that there is a set of multi-critical points that present anomalous behavior.  It is known that some multi-critical points may behave differently, giving rise to anomalous dynamical scaling properties and, as a consequence, different universality classes \cite{Multicritical 1, Multicritical 2}. The multi-critical points $\tilde{\lambda}^{(c)}$ , given by 
\begin{equation}
(\la_x,\la_y,h)=\left(\pm\frac{h-3}{2},\pm\frac{h+1}{2},h\right),\quad\forall\, h\in\mathbb{R},
\label{Multicritical line}
\end{equation}
have properties that are not present in other multi-critical points, like the point $(\la_x,\la_y,h)=(0,1,0)$ that was studied extensively in \cite{Stat Mech Cluster Ising}. In fact, 
the overlap between the ground state state corresponding to the critical points $\ket{\Omega(\la^{(c)}_i)}$ and the neighboring ground states, is very large. Moreover, also the overlap $F_1(\la_i')$ between such multi-critical ground states and  the subspace generated by excited pairs $\left\{\gamma^\dg_k\gamma^\dg_{-k} \ket{\Omega(\la^{(c)}_i)}\right\}_k$ is quite large:
\begin{align}
F_1(\la_i')&=\sum_{0\leq k\leq\pi}\left|\braket{\Omega(\la_i')|\gamma_k^\dg\gamma_{-k}^\dg|\Omega(\la_i^{(c)})}\right|^2. \label{Few excitations}
\end{align}
This overlap region will roughly follow the truncated surface given by Eq.(\ref{Cluster - Gapless 2}).

We can gain some physical intuition about these multi-critical if we rewrite the Hamiltonian as
\begin{align*}
H(h) = & \,3\left(\pm\sum_i \sig{i}{x}\sig{i+1}{x} + \sum_i \sig{i}{z}\right)\nonumber\\
& + \left(-\sum_i \sig{i-1}{x}\sig{i}{z}\sig{i+1}{x} \pm \sum_i \sig{i}{y}\sig{i+1}{y}\right)\nonumber\\
&+(h+3)\left(\pm\frac{1}{2}\sum_i (\sig{i}{x}\sig{i+1}{x} + \sig{i}{y}\sig{i+1}{y}) - \sum_i \sig{i}{z}\right).
\end{align*}
This corresponds to the sum of three critical Hamiltonians, namely a critical Ising model, a critical cluster-Ising model and a critical \textit{XX} model with a transverse field. Note that for $|h|\gg 1$, the \textit{XX} model term dominates. The ground state of this Hamiltonian corresponds, after a Jordan-Wigner transformation, to a completely empty (full) Fermi sea \cite{Sachdev-QPT}. This state is trivial from the entanglement point of view since it corresponds to a product state \cite{Entanglement spin chains}, but it affects the behavior of the fidelity susceptibility in its vicinity \cite{Multicritical fidelity 1, Multicritical fidelity 2}.

For the sake of concreteness we will concentrate on two of these points
\begin{align}
P1:\qquad (\la_x,\la_y,h)=&\left(-\frac{3}{2},\frac{1}{2},0\right),\\
P2:\qquad (\la_x,\la_y,h)=&\left(-2,0,-1\right).
\end{align}

\begin{figure}[]
\centering
\subfigure[]{
    \includegraphics[width=0.7\linewidth]{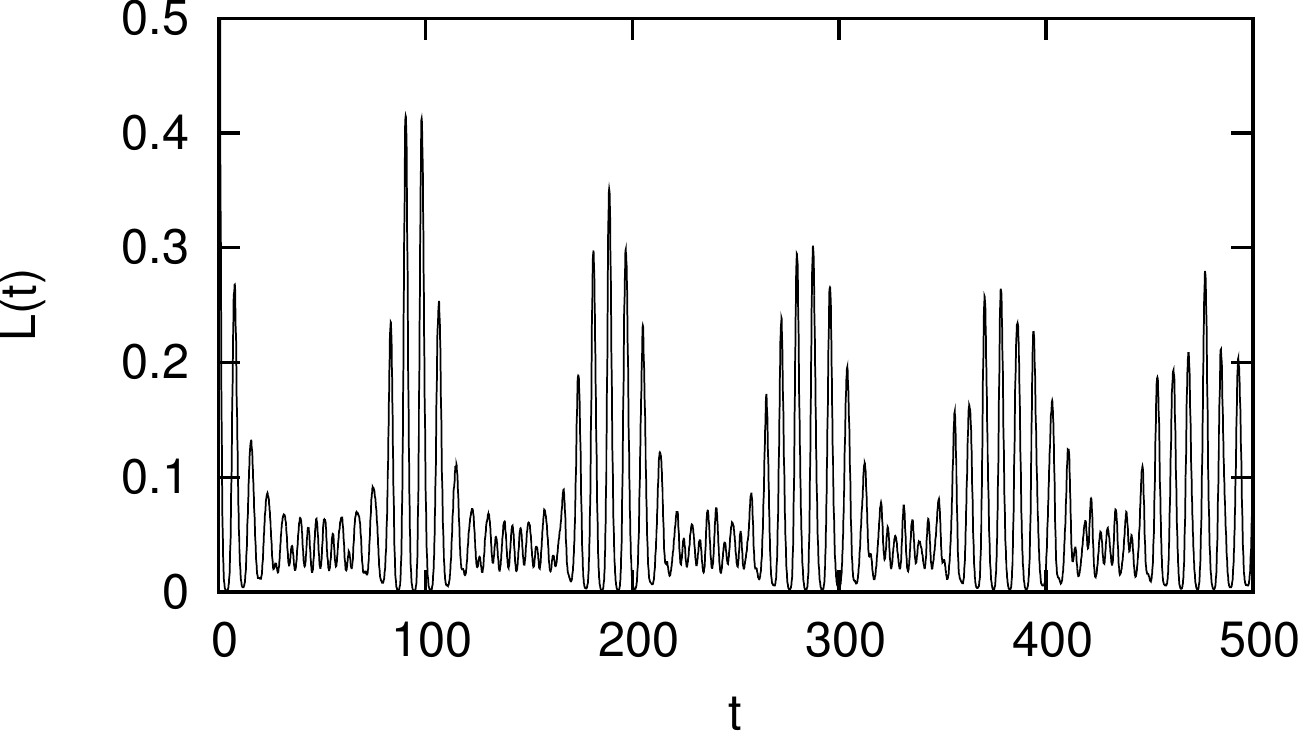}
}
\subfigure[]{
    \includegraphics[width=0.7\linewidth]{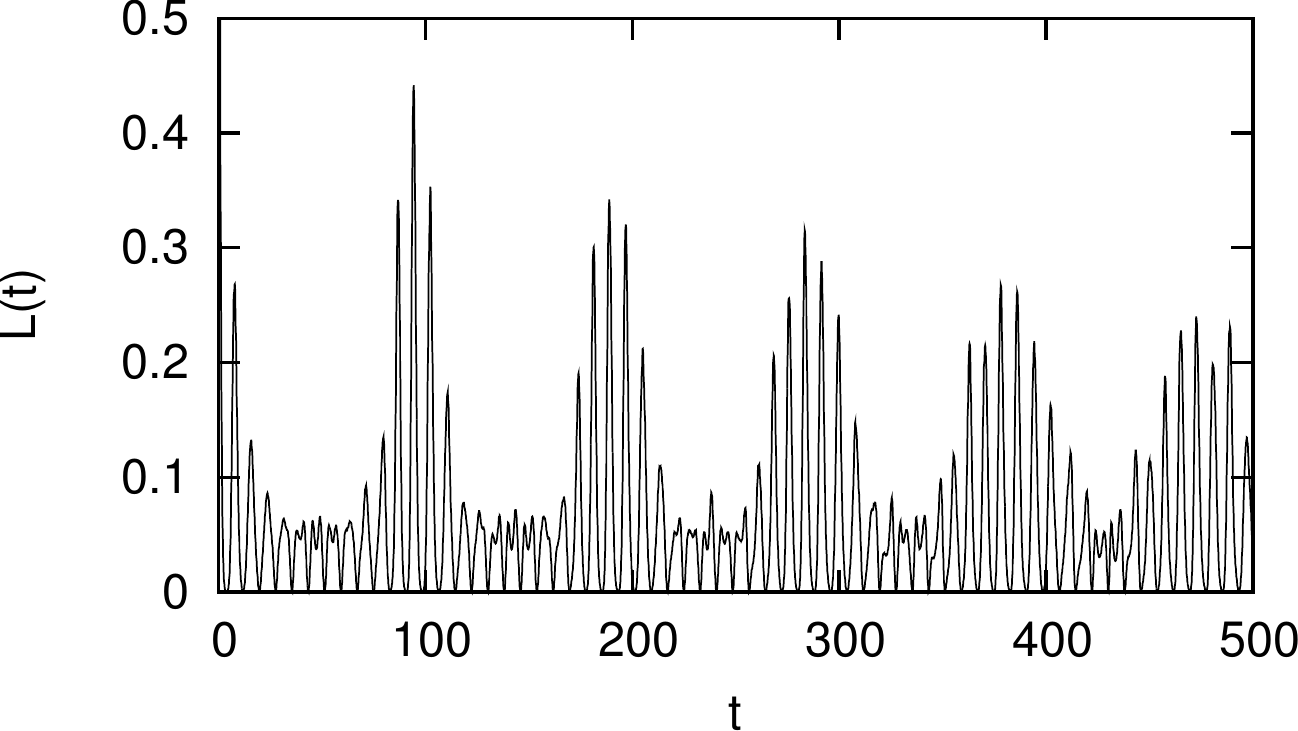}
}\\
\subfigure[]{
    \includegraphics[width=0.7\linewidth]{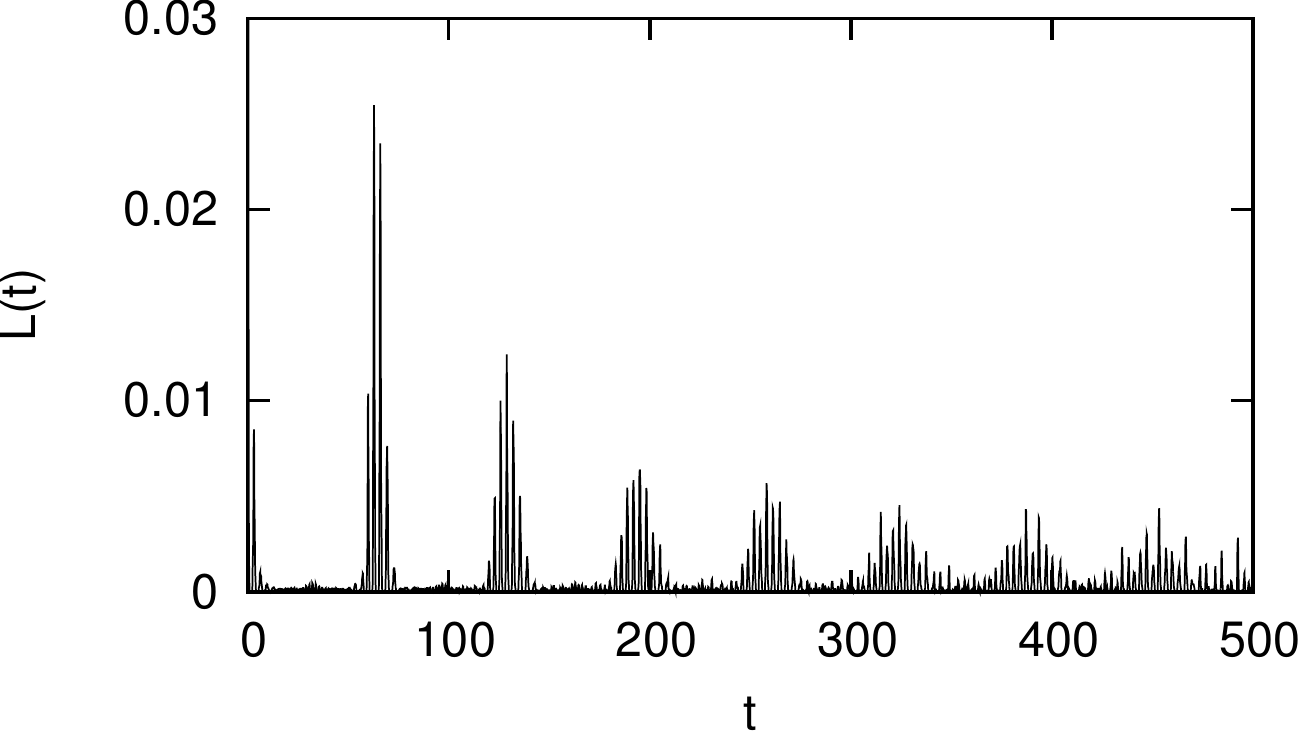}
}
\subfigure[]{
    \includegraphics[width=0.7\linewidth]{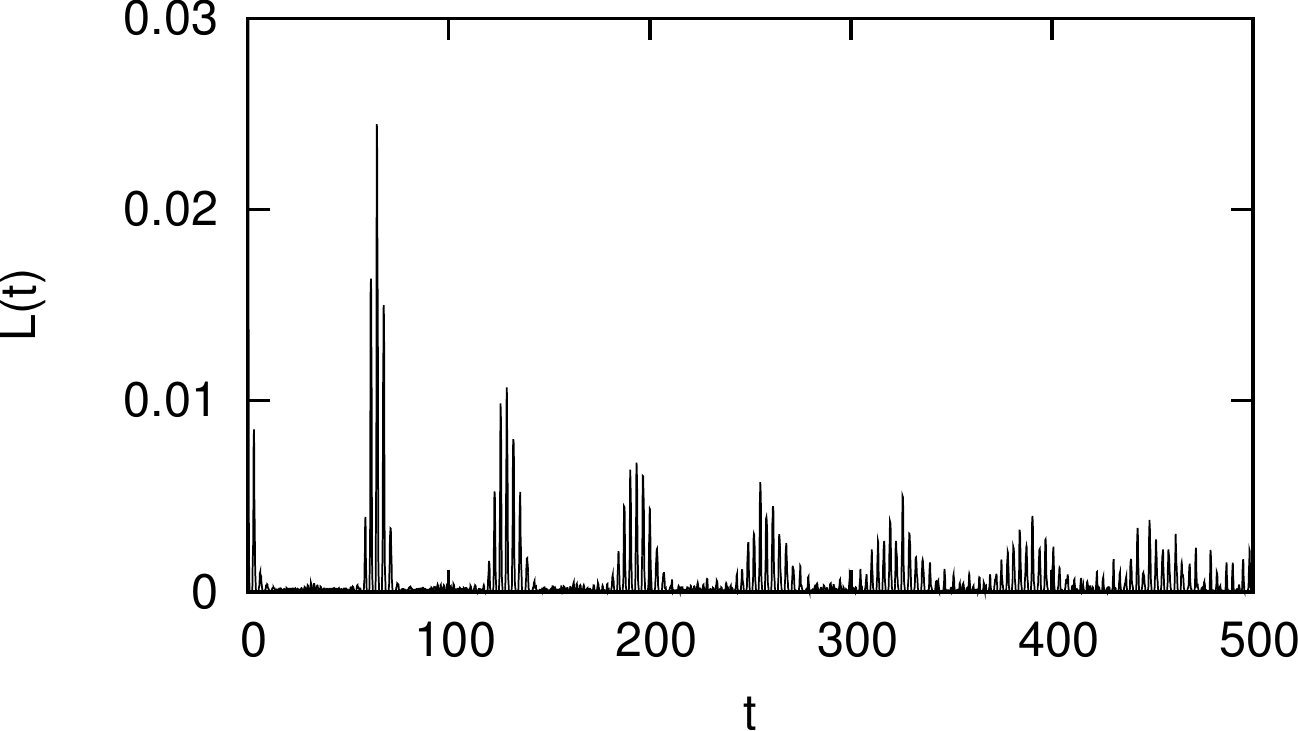}
}
    \caption{LE starting from the cluster phase to the critical point $\la_x=-\frac{3}{2}$, $\la_y=\frac{1}{2}$ and $h=0$ with $N=400$ starting from ($\la_y$ and $h$ kept fixed) (a) $\la_x=-1.3$, $q=1$, (b) $\la_x=-1.3$, $q=0$, (c) $\la_x=-1$, $q=1$, (d) $\la_x=-1$, $q=0$. }
   \label{LE QCP2 - Cluster}
\end{figure}

The overlap with the neighboring ground states is illustrated in figure \ref{h0-overlap} and \ref{h-1-overlap}. We see a significant overlap between the state at the quantum critical point and the neighboring ground states of the antiferromagnetic phase. This phenomenon is due to the fact that the perturbation corresponding to the parameter $\lambda_i$ is not sufficiently relevant \cite{Zanardi-Tensor}. The overlap $F_1(\la_i')$ \eqref{Few excitations} is also considered, showing that the most significant overlap is either with the ground state or just a few pairs of excitations. Note that $F_1$ is symmetric, so we also get the overlap between the subspace of a pair of excitations of the neighboring states and the critical point ground state.

The other critical points that do not belong to these critical lines  present a behavior similar to the one obtained for the \textit{XY} model in previous studies \cite{Zanardi - Ground State}. In those cases, the overlap with all the neighboring ground states decays very fast even for finite systems. In the present model, we expect this behavior for asymptotically large values of all the couplings $\la_x$, $\la_y$ and $h$ since the cluster interaction in the Hamiltonian \eqref{ClusterIsing Hamiltonian} becomes negligible in comparison and we obtain the usual \textit{XY} model. This fact implies that the two planes given by \eqref{Cluster - Gapless 1} correspond \emph{asymptotically} to the Ising critical lines and the hyperbolic surface \eqref{Cluster - Gapless 2} correspond \emph{asymptotically} to the \textit{XX} critical line. Note, however, that the universality class may be affected as we get close to the multicritical region we have been considering \cite{Multicritical 2}. 

\begin{figure}[]
\centering
\subfigure[]{
   \includegraphics[width=0.7\linewidth]{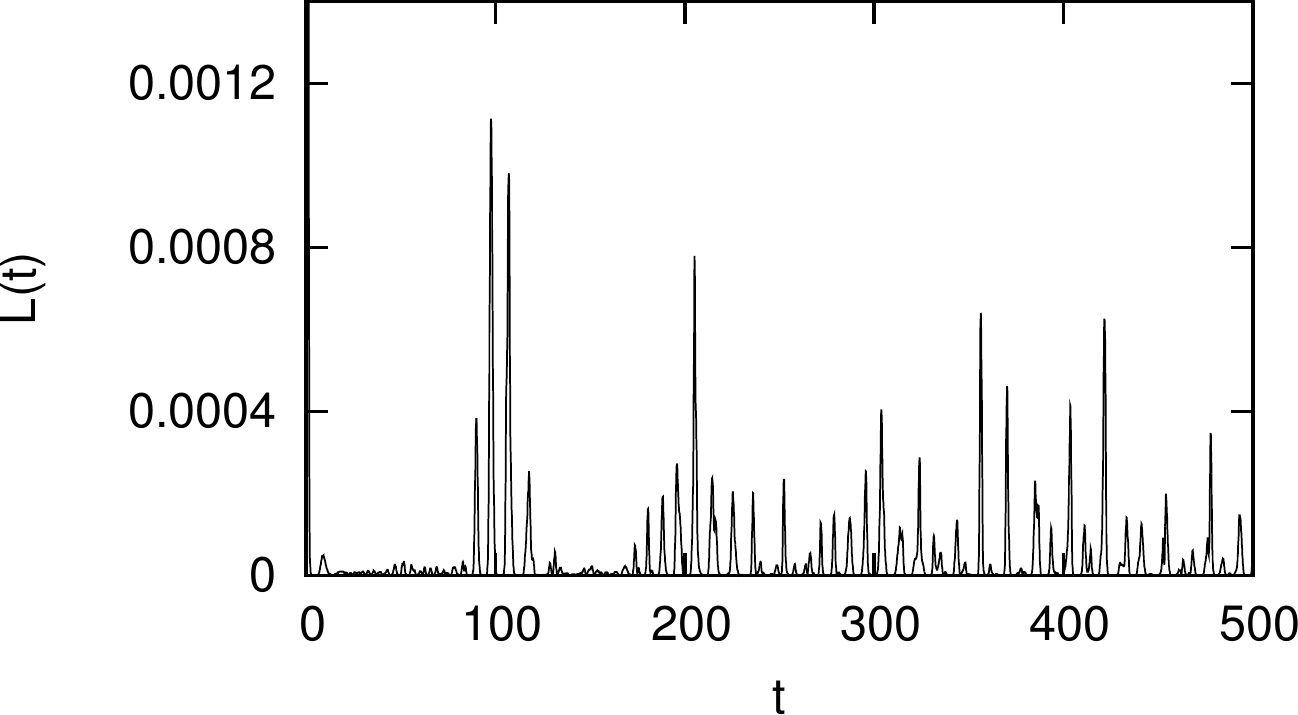}
\label{Polarized-LE}
}
\subfigure[]{
   \includegraphics[width=0.7\linewidth]{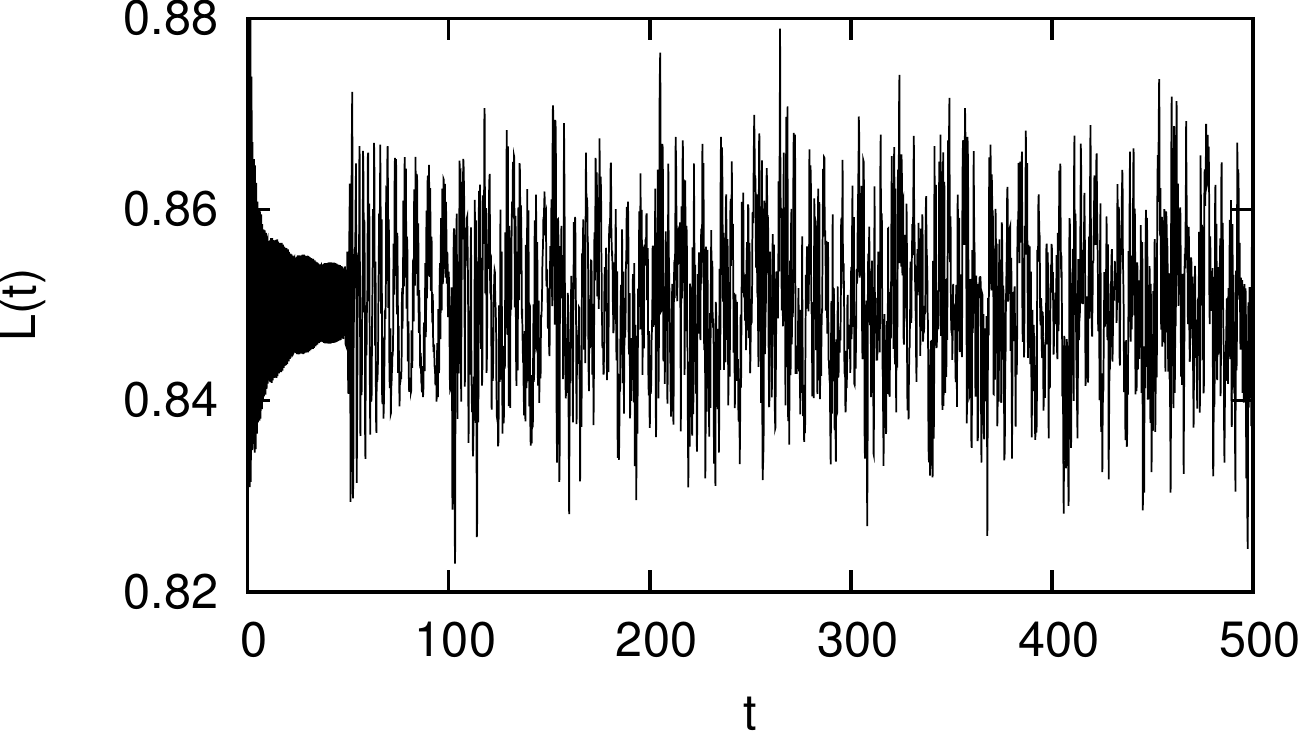}
\label{Ferromagnetic-LE}
}
    \caption{LE for the quenched cluster-Ising model to the critical point $\la_x=-\frac{3}{2}$, $\la_y=\frac{1}{2}$ and $h=0$ with $N=400$ starting from (a) $\la_y=0.7$, $q=1$ ($\la_x$ and $h$ kept fixed, polarized), (b) $\la_x=-1.7$, $q=1$ ($\la_y$ and $h$ kept fixed, ferromagnetic). }
    \label{LE QCP2 - Others}
\end{figure}

\textit{Quantum quenches and Loschmidt echo.---} At this point, we are ready to study the dynamics of the system after  a quantum quench. In order to quantify this we use the Loschmidt echo (LE). This quantity is widely used in many-body physics, in particular in the field of quantum chaos \cite{Loschmidt - Review, Loschmidt - Review 2, Zanardi - Unitary equilibration Loschmidt,Zanardi - Statistics Loschmidt echo}. Suppose we want to compare the dynamics under the Hamiltonians $H_1$ and $H_2$ (possibly time dependent) imposing the same initial conditions $\ket{\psi(t=0)}=\ket{\psi_0}$. In that case, we define the LE as
\begin{equation}
 \mathcal{L}(\psi_0,t)=\left|\braket{\psi_0|U_1(-t)U_2(t)|\psi_0}\right|^2,
\label{general LE}
\end{equation}
where $U_a(t)=\hat T\exp(-i\int_0^t H_a(t')dt')$ and $\hat T$ denotes time ordering. Note that $\mathcal{L}(t=0)=1$. In this paper, we will limit ourselves to ground states of one of the Hamiltonians, so that one of the unitaries in \eqref{general LE} acts trivially. This can be interpreted operationally as preparing the system in the ground state of the Hamiltonian with parameters $\la_i^{(1)}$, suddenly switching $\la_i^{(1)}\rightarrow \la_i^{(2)}$, and letting the system evolve with the new Hamiltonian. The LE reads
\begin{equation}
 \mathcal{L}(\la_i^{(1)}, \la_i^{(2)},t)=\left|\braket{\Omega(\la_i^{(1)})\left|U(t)\right|\Omega(\la_i^{(1)})}\right|^2.
\end{equation}
In this sense, the LE is a dynamical version of the ground state fidelity. 
High values of the LE mean that the system is approaching the initial state. Typically, the LE will decay exponentially at first and then start oscillating around a well-defined value \cite{Zanardi - Unitary equilibration Loschmidt,Zanardi - Statistics Loschmidt echo}. If the systems is finite, we expect the time evolution to be quasiperiodic, driving the system arbitrarily close to the initial state for long enough times. The system will experience \emph{revivals}, \textit{i.e.}, times where the value of the LE is greater than two standard deviations from the average. The structure of these revivals may be greatly affected by criticality \cite{Zanardi - Decay Loschmidt}.
The time evolution in the Cluster-\textit{XY} model after a quantum quench is given by
\begin{align}
 &\ket{\psi(t)}\nonumber\\
&=\prod_{0\leq k\leq\pi}\left(\cos(\frac{\chi_k}{2})+ie^{-i4t\Delta_k}\sin(\frac{\chi_k}{2})\gam_k^\dg \gam_{-k}^\dg\right)\ket{\Omega(\la_i^{(2)})}.
\end{align}
and the LE is thus
\begin{align}
\LE(t)&=|\braket{\psi(t)|\psi(0)}|^2\nonumber\\
&=\prod_{0\leq k\leq\pi}\left(1-\sin^2(\chi_k)\sin^2(2t\Delta_k)\right).
\label{LE spin chain}
\end{align}
In the following, we show the detailed analysis of the LE for different types of quenches. 
\begin{figure}[]
\centering
\subfigure[]{
    \includegraphics[width=0.7\linewidth]{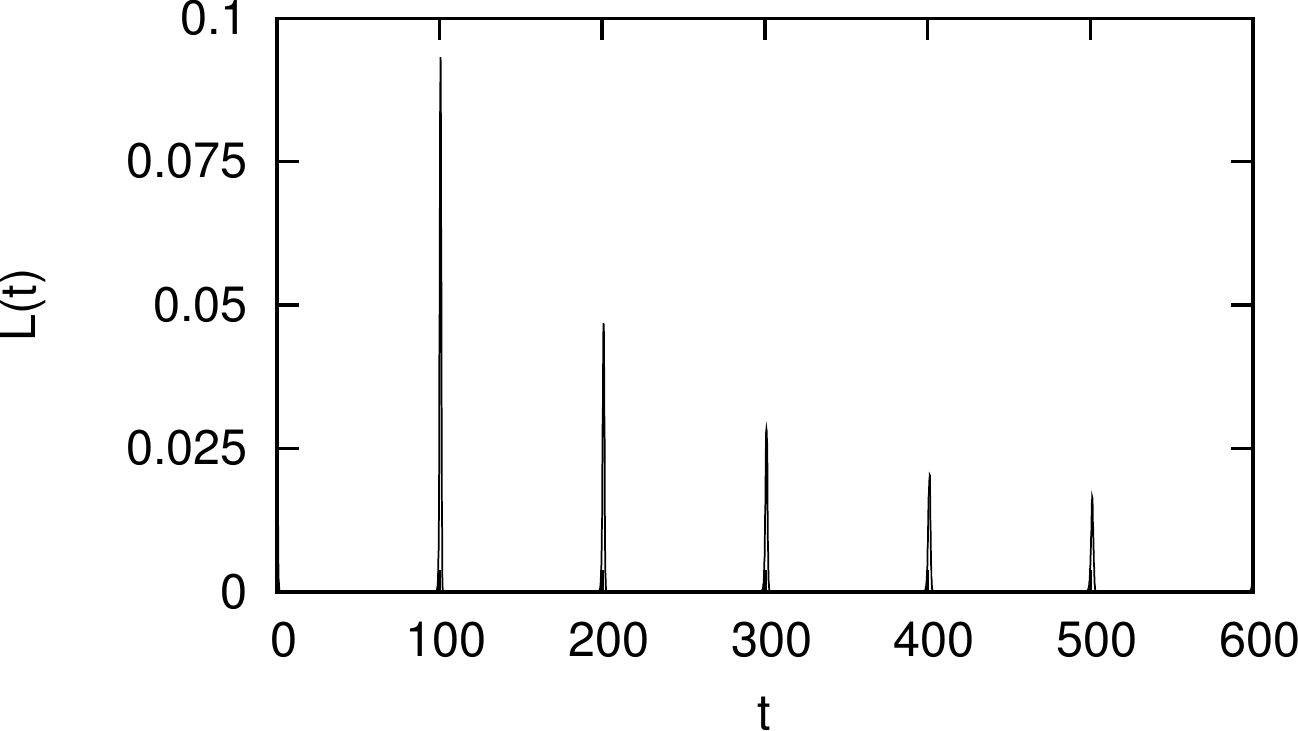}
}
\subfigure[]{
    \includegraphics[width=0.7\linewidth]{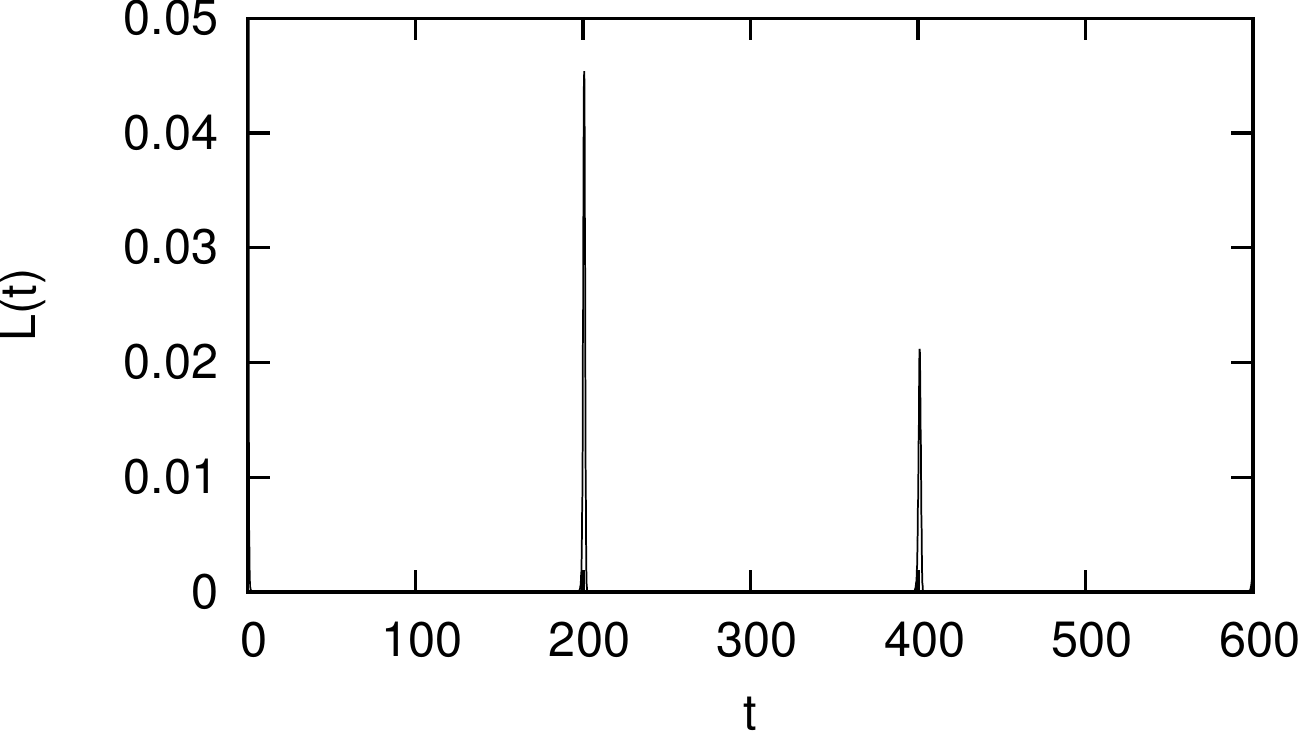}
}\\
\subfigure[]{
    \includegraphics[width=0.7\linewidth]{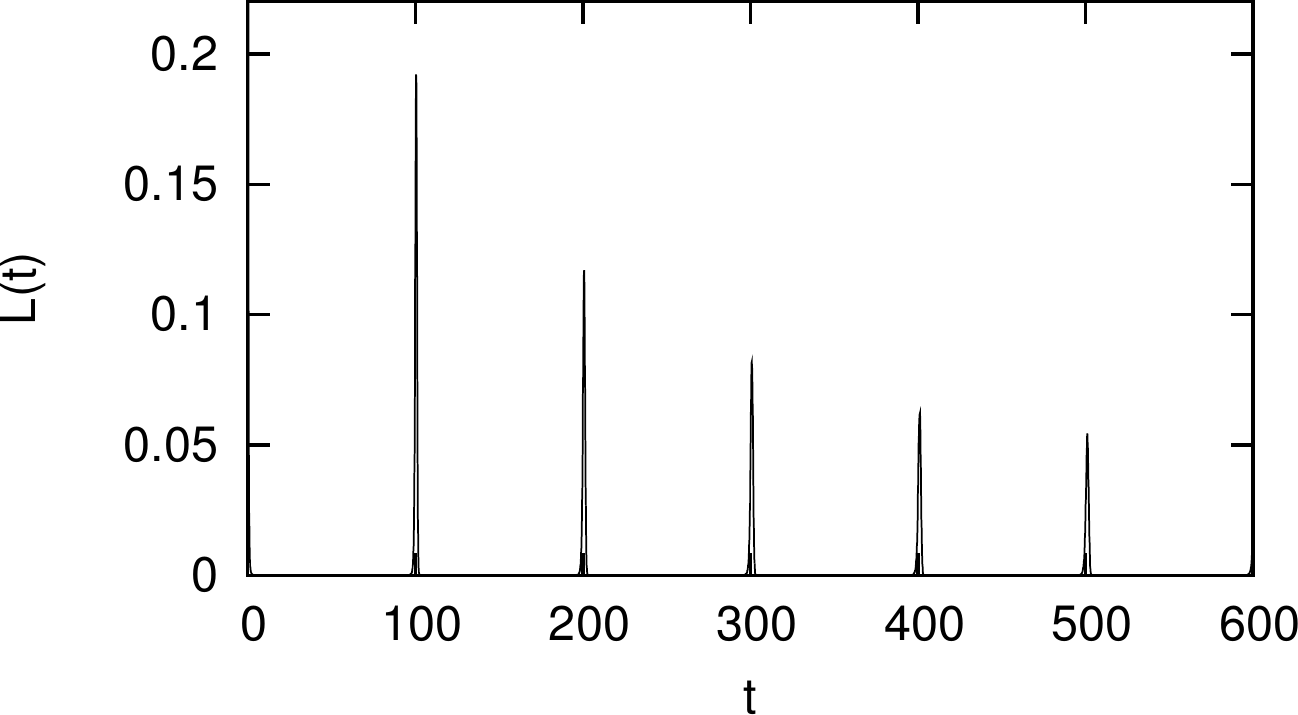}
}
\subfigure[]{
    \includegraphics[width=0.7\linewidth]{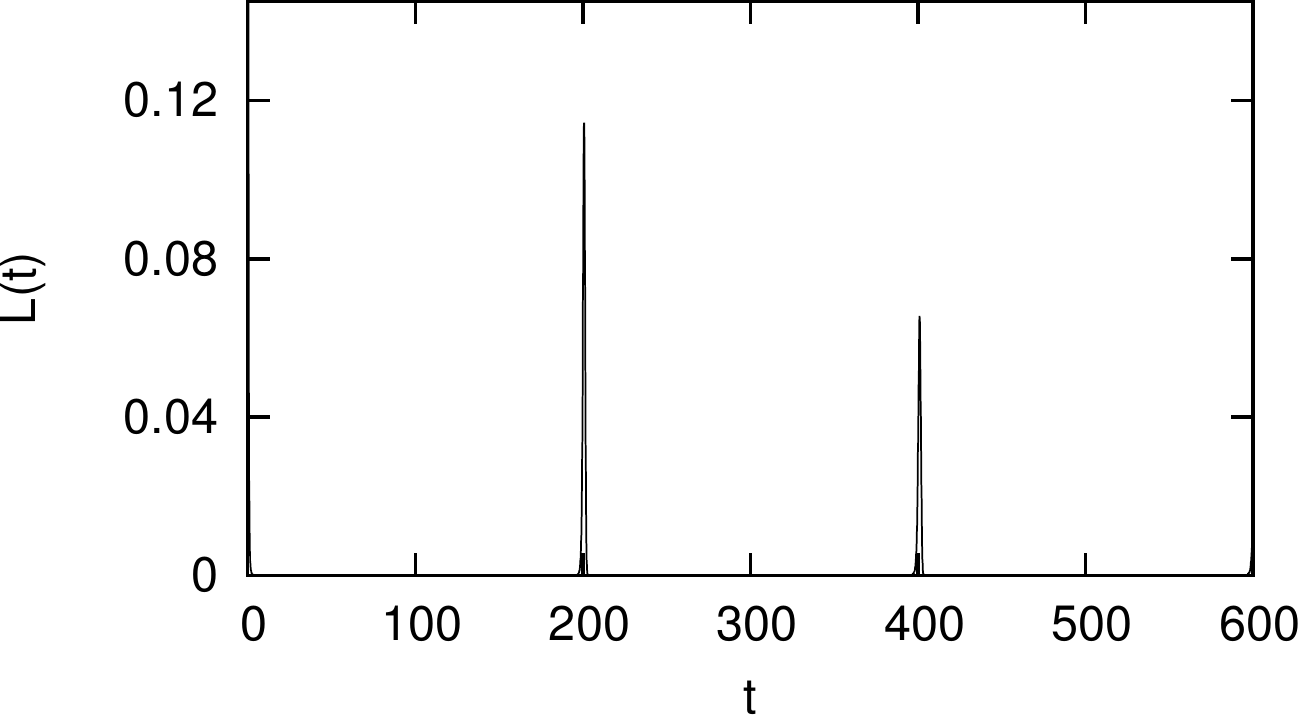}
}
    \caption{LE for the quenched cluster-Ising model to the critical point $\la_x=0$, $\la_y=1$ and $h=0$ with $N=400$ starting from ($\la_x$ and $h$ kept fixed) (a) $\la_y=0.8$, $q=1$ (cluster), (b) $\la_y=0.8$, $q=0$ (cluster), (c) $\la_y=1.2$, $q=1$ (antiferromagnetic), (d) $\la_y=1.2$, $q=0$ (antiferromagnetic). }
    \label{LE QCP1 - Cluster-Ising}
\end{figure}

Consider the critical point $h=0,\,\, \la_x=-\frac{3}{2},\,\, \la_y=\frac{1}{2}$. The critical Hamiltonian is
\[ H=-\sum_{i=1}^N \sig{i-1}{x}\sig{i}{z}\sig{i+1}{x}-\frac{3}{2}\sum_{i=1}^N \sig{i}{x}\sig{i+1}{x}+\frac{1}{2}\sum_{i=1}^N \sig{i}{y}\sig{i+1}{y}. \]
As we see in figure \ref{h0}, this point lies between three different phases. If we increase $\la_x$ ($\la_y$) we will be in the cluster (polarized) phase. Decreasing $\la_y$ or $\la_x$ results in a ferromagnetic state in the $x$ direction.

For this critical point the behavior of the LE depends on  which phase the system is prepared in. For the cluster state (Fig. \ref{LE QCP2 - Cluster}) the numerical simulations show that it will oscillate strongly away from the mean value. We also get a noticeable spreading of the revivals. If we start further from the critical point (\textit{i.e.} a stronger quench), the lines get closer, but we keep the insensitivity to the parity of the ground state ($Q=(-1)^q$). Note that the stronger the quench, the sooner the revivals will happen. If we start from the polarized phase (Fig. \ref{Polarized-LE}), the behavior of the LE is somewhat different. Once again, the oscillations make the revival structure insensitive to the parity sector.

On the other hand, starting from the ferromagnetic phase changes the LE completely (Fig. \ref{Ferromagnetic-LE}). The numerical simulations show that it will oscillate randomly around a relatively high mean value. There is no outstanding structure for the revivals as we had in the previous quenches. Increasing the size of the quench will give us basically the same result, only decreasing the mean value. This is consistent with the significant overlap of the ground state of this critical point with the neighboring ferromagnetic ground states (Fig. \ref{P1-gs}).


Consider now $h=0,\,\, \la_x=0,\,\, \la_y=1$. For these values, the critical Hamiltonian is
\[ H=-\sum_{i=1}^N \sig{i-1}{x}\sig{i}{z}\sig{i+1}{x}+\sum_{i=1}^N \sig{i}{y}\sig{i+1}{y}. \]

This critical point lies on the interface of four different phases (Fig. \ref{h0}). If $\la_y$ is increased, we will be in a region that can be connected to an antiferromagnet in the $y$ direction. If $\la_y$ is decreased, we will be in the cluster region. A positive $\la_x$ will turn the system into an antiferromagnet in the $x$ direction, while a small negative $\la_x$ will put it in a region that can be connected to a separable state polarized in the $z$ direction. 

\begin{figure}[]
\centering
\subfigure[]{
    \includegraphics[width=0.7\linewidth]{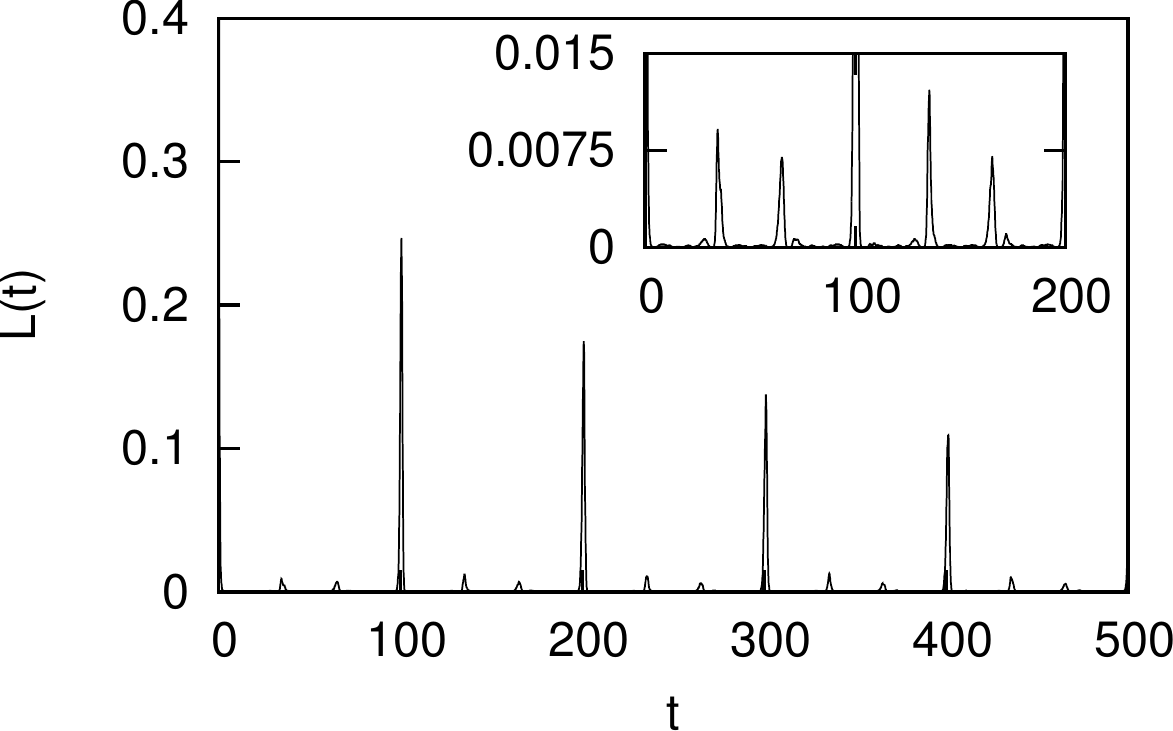}
}
\subfigure[]{
    \includegraphics[width=0.7\linewidth]{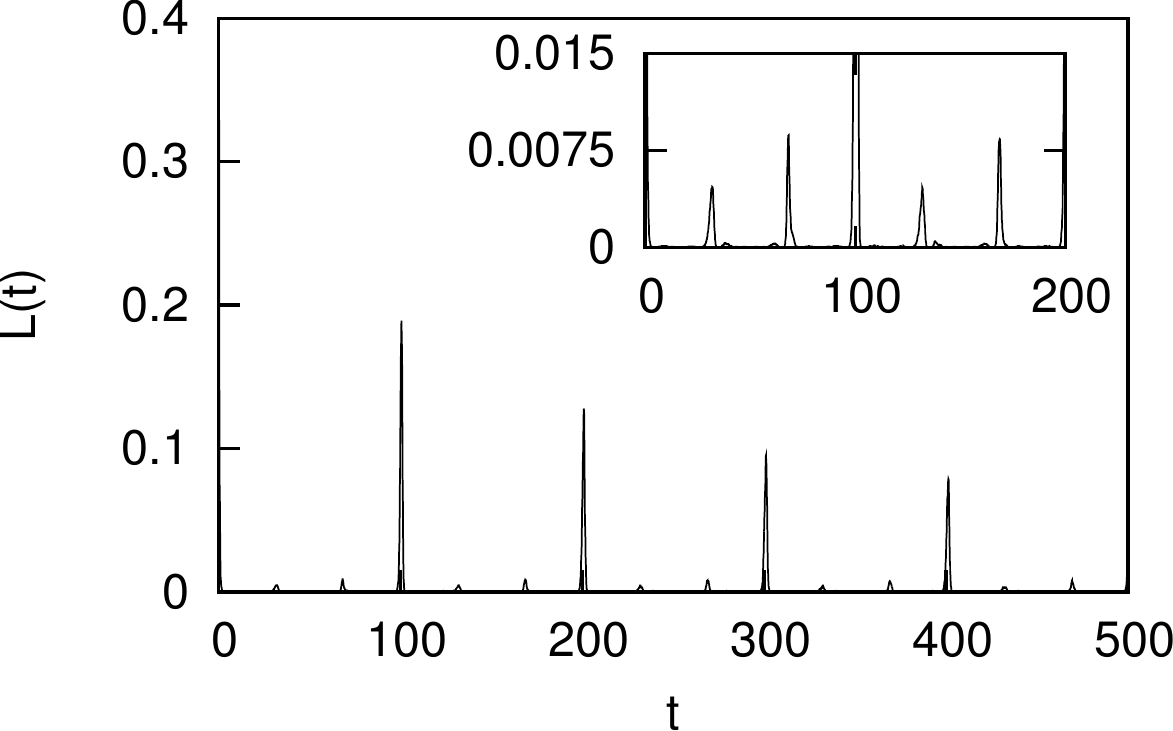}
}
    \caption{LE for the quenched cluster-Ising model to the critical point $\la_x=0$, $\la_y=1$ and $h=0$ with $N=400$ starting from ($\la_y$ and $h$ kept fixed) (a) $\la_x=-0.2$, $q=1$, (b) $\la_x=0.2$, $q=1$. }
    \label{LE QCP1 - 2 - Cluster-Ising}
\end{figure}

\begin{figure}[h]
\centering
\subfigure[]{
    \includegraphics[width=0.7\linewidth]{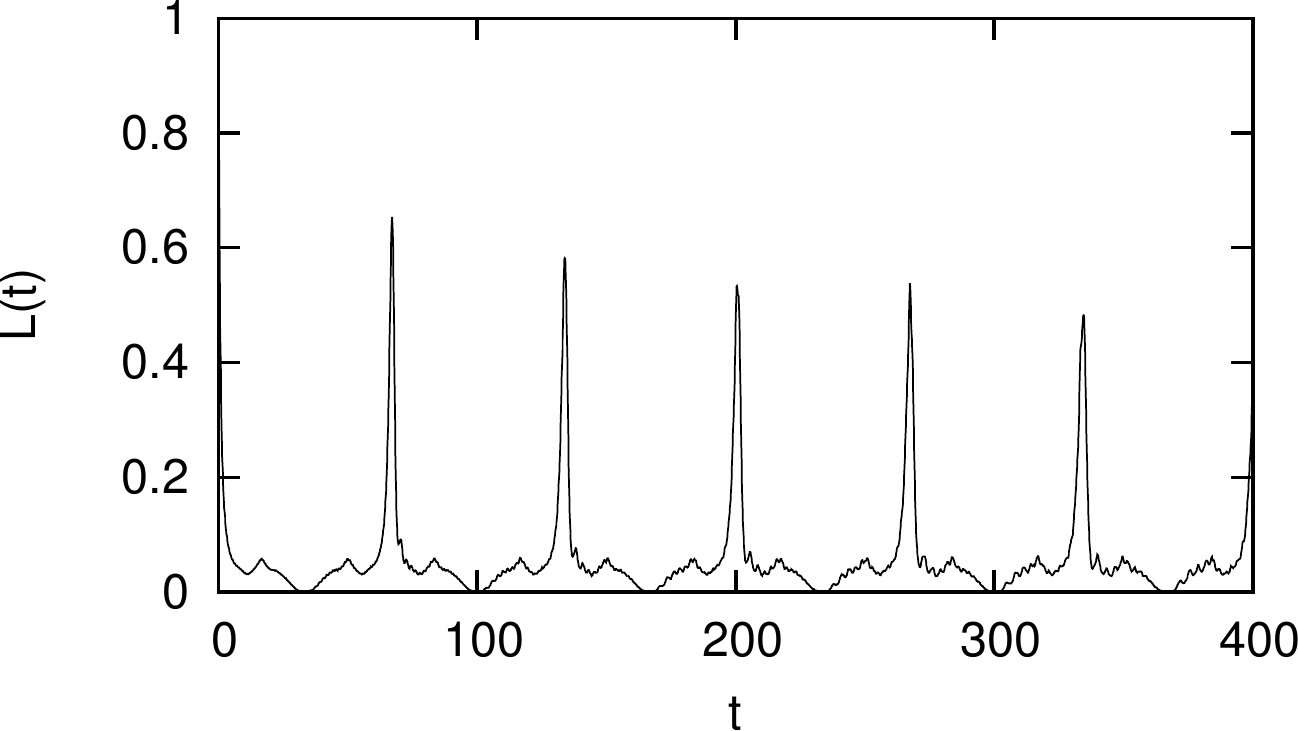}
}
\subfigure[]{
    \includegraphics[width=0.7\linewidth]{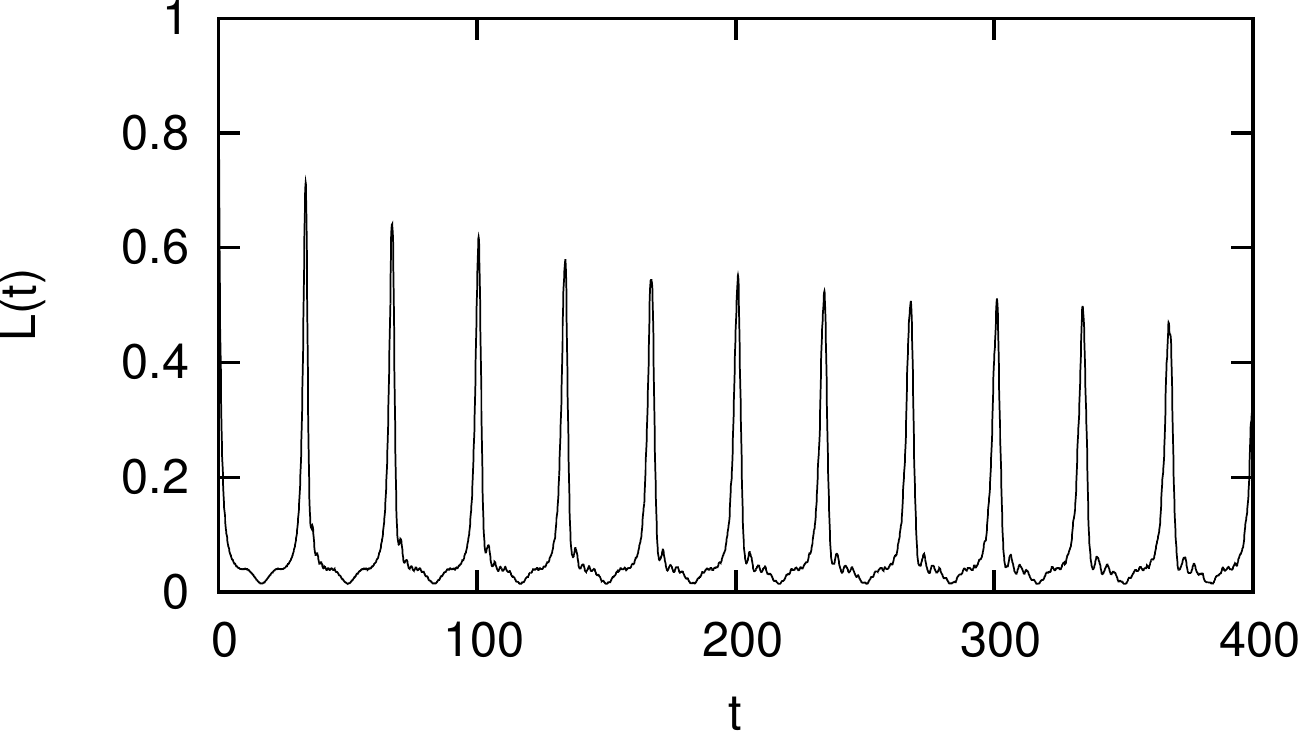}
}
\subfigure[]{
    \includegraphics[width=0.45\linewidth]{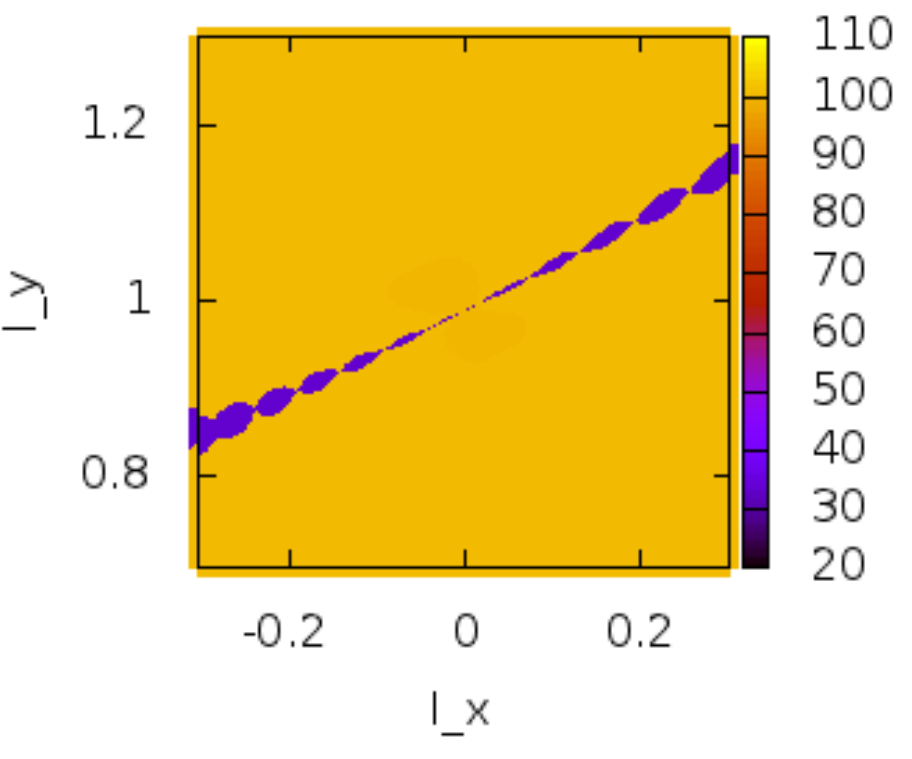}
\label{Revival map}
}
\subfigure[]{
    \includegraphics[width=0.45\linewidth]{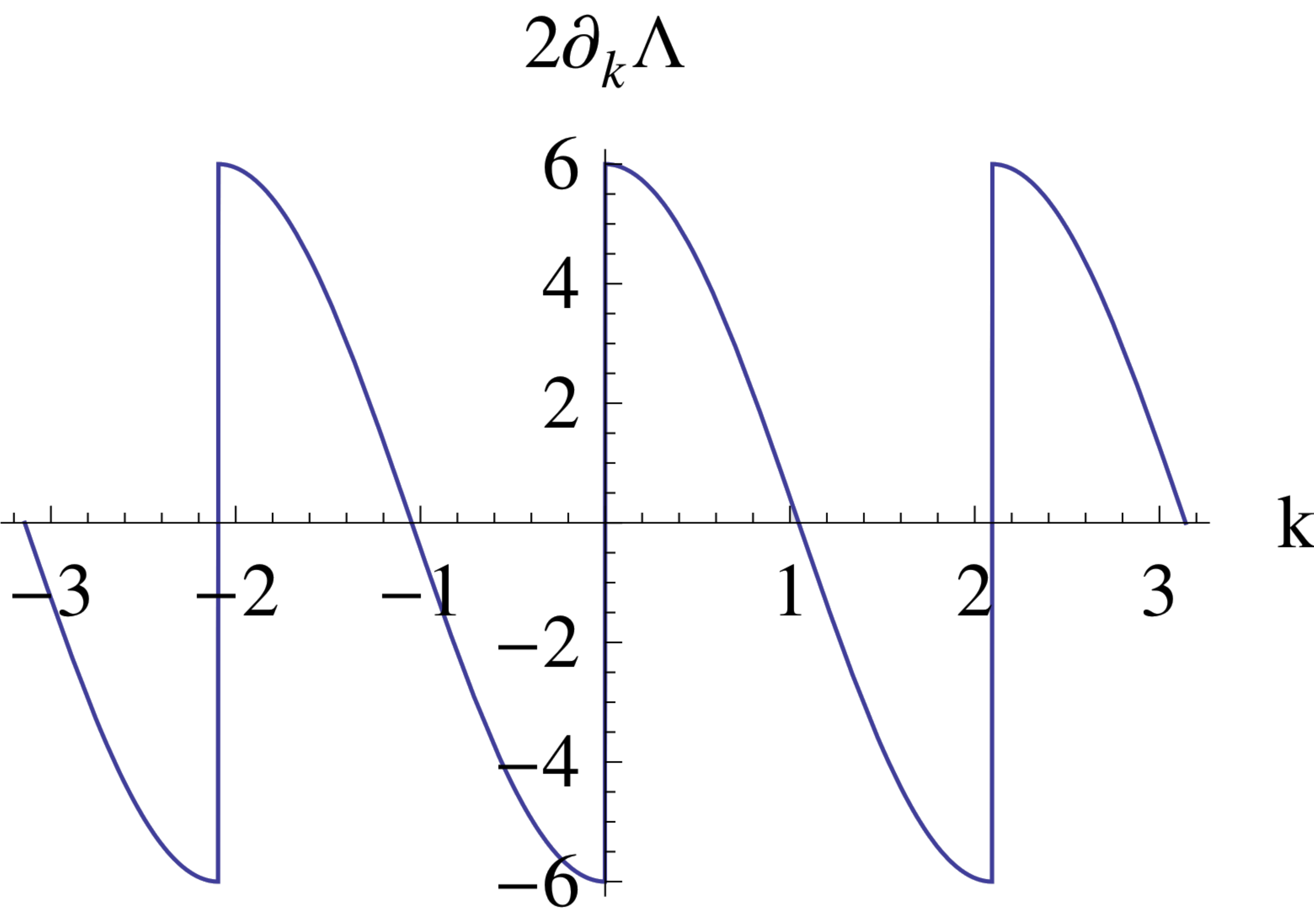}
\label{Group velocity}
}
    \caption{(Color online) LE for the quench to the critical point $\la_x=0$, $\la_y=1$ and $h=0$ along the critical line $\la_y^2-\la_x\la_y-1=0$ with $N=400$ starting from ($\lambda_y$  along the critical line and  $h=0$ kept fixed ) (a) $\la_x=0.1$, $q=0$. Notice that the revivals with odd parity are destroyed by interference (b) $\la_x=0.1$, $q=1$. (c) Revival time for critical quenches to the same critical point starting in the neighboring ground states. Notice the critical line is detected by the revival time. (d) Group velocity $2\pd_k\Delta_k$ for the critical point. The maximum value is $2\pd_k\Delta_k|_{\max}=6$.}
    \label{LE-Critical parabola}
\end{figure}

The behavior of the LE after a quench of $\la_y$ (Fig. \ref{LE QCP1 - Cluster-Ising}) is similar to the one obtained for critical quenches in the \textit{XY} model  \cite{RevivalLiebRobinson}. It features the same sensitivity to the parity that cancels the odd revivals for even parity ($q=0$). This can be understood easily by noting that for $k=\pi/N$ and $t=\frac{N}{4}\left|\frac{\pd\Delta_k}{\pd k}\right|_\text{max}^{-1}$, we get $\chi_k\sim \frac{\pi}{2}$ (as can be seen in the fidelity) and $2t\Delta_k\sim \frac{\pi}{2}$, canceling the LE (eq. \eqref{LE spin chain}).

Starting with a small non-zero value for $\la_x$ and quenching the system to this critical point, the LE will behave roughly in the same way (Fig. \ref{LE QCP1 - 2 - Cluster-Ising}). However, we get two small peaks before the first revival. Strictly speaking, they will not be revivals according to the definition because they are less than two standard deviations away from the mean value of the LE. We can understand them as dynamical responses given by the Bogoliubov quasiparticles. 

We can also quench to the neighboring critical points along the critical curve $\la_y^2-\la_x\la_y-1=0$, see  (Fig. \ref{LE-Critical parabola}). This means that both the starting and quenching Hamiltionian lie on this curve. In this case, the main difference will be the revival time, that will be exactly one third of the one found in the previous quenches. Note that we do not get this phenomenon for the critical line $\la_y=-\la_x+1$, $h=0$.

In  \cite{RevivalLiebRobinson}, the phenomenon of the revivals after a quantum quench is interpreted as a recombination of the fastest quasiparticles in the system. In general, the fastest excitations in the system have a speed that is upper bounded by the Lieb-Robinson speed $v_\text{LR}$ \cite{lrb, Lieb-Robinson speed, Lieb-Robinson thesis}. This upper bound gives a lower bound to the revival time $T_{\text{rev}}\gtrsim\frac{N}{2v_\text{LR}}$. Following \cite{lrs}, we find $v_{LR}\simeq 3.2 e/\sqrt{2}=6.15$ for the critical point $\la_y=1$, $\la_x = 0$ and $h=0$ and therefore the maximum speed of quasi particles given by the maximum group velocity $2\pd_k\Delta_k|_{\max}=6$ (Fig. \ref{Group velocity}) is compatible with the Lieb-Robinson bound.


\textit{Conclusions.---} In this paper, we studied the phase diagram and quench behavior of the Cluster-\textit{XY} model, a spin chain where the usual \textit{XY} interactions in a transverse field are competing with a cluster three-body term. This model also describes the effective behavior of the edge in a 2D fermionic symmetry protected topological state with Z2 symmetry \cite{edge}. The Cluster-\textit{XY} model is exactly solvable by standard techniques, and the study has been conducted using the tools of fidelity susceptibility and Loschmidt Echo (LE).  This model, inspired by proposed implementations of quantum computation, provides a new benchmark with an interesting phenomenology and a much richer phase space coming from the competition of the different interactions. We were able to characterize the critical regions and the distribution of phases using the quantum geometric tensor. We found that the phase diagram is completely characterized by the fidelity. Noteworthy, the ground state of some of the critical points present a large overlap with the ground state and few-excitations subspaces of neighboring non-critical regions. The behavior away from equilibrium is also non trivial. We showed that different critical points have qualitatively different effects on the LE. The long-time structure and the revival times of the LE depend on the initial phase of the quantum quench and the final critical point. This provides further phenomenology for the study of generic responses to critical quantum quenches.

In  \cite{ClusterAFM, won}, it was shown that the Cluster-Ising model with open boundary conditions has a four fold degenerate ground space, which possesses symmetry protected topological order \cite{symmprot,symp, symp2}. We expect that this model has similar features \cite{next}, though the presence of the non trivial phases between the cluster phase and the ferromagnetic phase makes the situation more complicated. A promising route to the characterization of topological orders is the study of their entanglement spectrum \cite{entspec1, entspec2}. In a symmetry protected one-dimensional spin-one chain in the Haldane phase, the topological order is revealed in a double degeneracy of the entanglement spectrum \cite{entspec3}. It would hence be interesting to study the entanglement spectrum of the cluster-\textit{XY} model, to gain more insight into the properties of the symmetry protected topological order. 
In particular, it would be interesting to study the robustness of the information encoded in such ground space after a quantum quench, and whether the quench breaks or not the symmetry that protects the topological order. Finally, it would be interesting to study the model in presence of disorder. In order to obtain reliable results, more sophisticated numerical techniques may be used such as matrix product states.

\textit{Acknowledgments.---} S.M. would like to thank Felix Flicker, Holger Haas, Guifr{\' e} Vidal and Marek Rams for useful discussions.  Research at Perimeter Institute for Theoretical
Physics is supported in part by the Government of Canada through NSERC and
by the Province of Ontario through MRI.

\end{document}